\documentclass[iop,revtex4]{emulateapj}

\newcommand{\cha}{\textit{Chandra}}

\def\flu{{ erg s$^{-1}$} cm$^{-2}$}
\def\nustar{{\it NuSTAR}}
\def\swi{{{\it Swift}-BAT}\/}
\usepackage{natbib}
\usepackage{color}
\usepackage{graphicx}
\usepackage{gensymb}
\usepackage{enumitem}
\usepackage[toc,page]{appendix}
\usepackage{pdflscape}
\usepackage{rotating}

\begin{document}


\title{\textit{Chandra} and \textit{NuSTAR} follow-up observations of \textit{Swift}-BAT selected AGN}


\author{S. Marchesi\altaffilmark{1}, L. Tremblay\altaffilmark{1}, M. Ajello\altaffilmark{1}, L. Marcotulli\altaffilmark{1}, \\ A. Paggi\altaffilmark{2}, G. Cusumano\altaffilmark{3}, V. La Parola\altaffilmark{3}, A. Segreto\altaffilmark{3}}

\altaffiltext{1}{Department of Physics and Astronomy, Clemson University, Clemson, SC 29634, USA}
\altaffiltext{2}{Harvard--Smithsonian Center for Astrophysics, 60 Garden Street, Cambridge, MA 02138, USA}
\altaffiltext{3}{INAF - Istituto di Astrofisica Spaziale e Fisica Cosmica, Via U. La Malfa 153, I-90146 Palermo, Italy}

\begin{abstract}
Based on current models of the Cosmic X-ray Background (CXB), heavily obscured Active Galactic Nuclei (AGN)
are expected to make up $\sim 10\%$ of the peak emission of the CXB and $\sim 20\%$ of the total population of AGN, yet few of these sources have been recorded and characterized in current surveys.
Here we present the \cha\ follow-up observation of 14 AGN detected by \swi. 
For five sources in the sample, \nustar\ observations in the 3--80\,keV band are also available. The X-ray spectral fitting over the 0.3--150\,keV energy range allows us to determine the main X-ray spectral parameters, such as the photon index and the intrinsic absorption, of these objects, and to make hypotheses on the physical structures responsible for the observed spectra. We find that 13 of the 14 objects are absorbed AGN, and one is a candidate Compton thick AGN, having intrinsic absorption N$_{\rm H}>10^{24}$\,cm$^{-2}$.
Finally, we verified that the use of \nustar\ observations is strategic to strongly constrain the properties of obscured AGN, since the best-fit values we obtained for parameters such as the power-law photon index $\Gamma$ and the intrinsic absorption $N_{\rm H}$ changed sometimes significantly fitting the spectra with and without the  use of \nustar\ data.

\end{abstract}

\keywords{galaxies: active -- galaxies: nuclei --X-rays: galaxies}

\section{Introduction}

Active Galactic Nuclei (AGN, both of the absorbed and unabsorbed type) can explain most of the Cosmic X-ray Background (CXB) emission, i.e., the diffused X-ray emission observed in the 1 to $\sim$200--300\,keV energy range \citep{alexander03,gilli07,treister09}. Below 10\,keV the CXB emission has been almost entirely resolved using 0.5--10\,keV instruments like \cha\ \citep[see, e.g.,][]{cappelluti17}. In this energy range, the CXB emission is predominantly produced by unobscured (i.e., having absorbing column density $N_{\rm H}$$<10^{22}$\,atoms cm$^{-2}$) or Compton thin (10$^{22}<N_{\rm H}<10^{24}$\,cm$^{-2}$) AGN.
However, moving towards higher energies the scenario changes significantly: in fact, only $\sim$30\,\% of the CXB peak emission \citep[at $\sim$30\,keV][]{ajello08a} has been directly resolved \citep{aird15,civano15,mullaney15,harrison16}, using the Nuclear Spectroscopic Telescope Array \citep[\nustar][]{harrison13}.
Furthermore, the observed fraction of the so-called Compton thick (CT, N$_{\rm H}>10^{24}$\,cm$^{-2}$) AGN \citep[$f_{\rm CT}\sim$5-10\%, 
see, e.g.,][]{comastri04,dellaceca08,burlon11,vasudevan13,ricci15}, that are predicted to be significant contributors to the CXB peak, is still below the model predictions \citep[$\sim$10-35\,\%, see, e.g.,][]{gilli07,treister09}.

The discrepancy between the models predictions and the observed fraction of CT-AGN is most likely due to an observational bias in current surveys. Indeed, in heavily obscured sources Compton scattering and absorption can significantly reduce the observed source flux even above $\sim$10\,keV\citep{burlon11}, making CT-AGN very hard to detect.
For column densities $\leq$10$^{25}$\,cm$^{-2}$ part of the direct nuclear emission pierces through the torus at energies $\geq10$\,keV, while for larger column densities only the scattered, indirect component is visible \citep{matt99,murphy09}. Consequently, observational biases must be properly taken into account when the intrinsic CT-AGN fraction is computed. For example, \citet{burlon11} analyzed a complete sample of $\sim$200 local (i.e. z$<$0.1) AGN detected by {\it Swift}-BAT in the 15-55\,keV band and measured an observed fraction of Compton-thick AGN (relative to the whole population) $f_{CT}\sim$5\,\%. However, several models developed to characterize the heavily obscured AGN emission \citep[e.g.,][]{murphy09,brightman11} clearly show that even above 10\,keV the AGN emission is strongly suppressed by absorption and Compton scattering, therefore biasing X-ray surveys against the detection of heavily obscured AGN. After carefully modeling this additional selection effect, \citet{burlon11} recovered the intrinsic absorbing column density distribution, showing that indeed CT-AGN are fairly numerous, representing at least $\sim20$\,\% of the whole population.
To properly measure this observational bias, and therefore estimate the intrinsic CT-AGN population, one requires large, complete samples of nearby AGN, detected in both the 0.3--10\,keV and above 10\,keV, to properly measure X-ray spectral parameters such as the photon index $\Gamma$ and the intrinsic absorption N$_{\rm H}$.

Moreover, absorption within the torus is probably not due to a continuous distribution of gas and dust, being instead likely due to a clumpy distribution of optically thick, dusty clouds \citep{elitzur06,honig07,nenkova08}. This belief is strengthened by the discoveries of the occultation of the X-ray nucleus of NGC 1365 by a Compton-thick cloud \citep{risaliti07}. Nonetheless, at the present day only small numbers of CT-AGN have been studied with the high-quality data which is required to characterize them properly, mostly thanks to \nustar, the first telescope with focusing optics at $>$10 keV \citep[see][among others]{balokovic14,puccetti14,brightman15,rivers15,masini16,puccetti16}. Thus, to probe directly the physics of the torus and of the accretion disk one needs to build large samples of 2--100\,keV spectra of heavily obscured AGN.

In this work, we analyze the 0.3-150 keV spectrum of 14 AGN detected in the 60 month \textit{Swift}-BAT (15-55 keV) survey \citep{ajello12} and for which we obtained $\sim$5\,ks follow-up observation with Chandra ACIS-I. These objects were originally selected either because their properties (either their Seyfert 2 optical type or the absence of a bright ROSAT counterpart) suggest the presence substantial intrinsic absorption along the line of sight, or because they were the last few sources of the 60 month survey without 0.3--7\,keV coverage. We note that these sources are all detected in the Palermo 100-month BAT catalog (Segreto et al. in prep.), which we seek to make complete in terms of X-ray spectral charactherization.
Moreover, five sources also have been observed with \nustar\ in the 3-80\,keV band. Combining the observations from the different telescopes, we aim to properly characterize the main physical features of these sources, a fraction of which are expected to be CT-AGN. Thus, in the following, we utilize appropriate models to characterize the different features commonly observed in AGN X-ray spectra and seek to present an appropriate physical interpretation to the best-fit models that we obtain.

In the sections below we will address the methods and procedures used throughout our work. In Section \ref{sec:data_red} we describe the data reduction procedure. Section \ref{sec:analysis} covers the steps taken to develop appropriate best-fit models for each source spectrum and the physical implications of the individual components of these models. Finally, our results are discussed and summarized  in Section \ref{sec:discuss}.

\begin{table*}
\centering
\scalebox{0.95}{
\begin{tabular}{lcclcccccc}
\hline
\hline
SWIFT Name  & Source Name & R.A. & Decl.  & Type &  Redshift &  Obs. Date & Obs ID  & Exp Time & $d_{\rm Cha-BAT}$\\ 
4PBC & &deg & deg & & &D-M-Y &&Ks & arcmin\\
(1) & (2) & (3) & (4) & (5) & (6) & (7) & (8) & (9) & (10)\\
\hline
J0231.9$-$3639  & IC 1816 & 37.9625 & -36.6721 &  Sy2 & 0.017  & 31-08-2012 & 00014035 & 5.02 & 0.82\\
J0233.4+2758  & Mrk 1179 & 38.3431 & 27.9369 & Sy1 & 0.038  & 21-06-2012 & 00014036 & 5.02 & 2.56\\
J0251.6$-$1640  & NGC 1125 & 42.9178 & -16.6510 & Sy2 & 0.011  & 10-10-2012 & 00014037 & 5.02 & 1.07\\
J0455.8$-$7531 & ESO 33$-$2 & 73.9957 & -75.5412 & Sy2 & 0.018  & 07-10-2014 & 00016156 & 4.91 & 0.96\\
J0543.5$-$2738  & ESO 424$-$12 & 85.8873 & -27.6514 & Sy2 & 0.010  & 20-06-2012 & 00014039 & 5.02 & 1.01\\
J0640.6$-$4320  & 2MASXJ06403799$-$4321211 & 100.1583 & -43.3558 & G & 0.061  & 27-08-2012 & 00014040 & 5.02 & 1.11\\
J0920.0+3711  & IC 2461 & 139.9918 & 37.1913 & Sy2 & 0.008  & 16-04-2014 & 00016157 & 5.11 & 0.78\\
J1214.2+2932  & Was 49 & 183.6065 & 29.6035 & Sy2 & 0.061  & 25-03-2012 & 00014042 & 5.02 & 1.12\\ 
J1339.7+5548  & 2MASSJ13393397+5546142 & 204.8916 & 55.7706 & Sy1 & 0.123  & 19-02-2012 & 00014043 & 5.02 & 2.85\\
J1353.5$-$1125  & 2MASXJ13532820$-$1123055 & 208.3675 & -11.3850 & G & 0.069  & 27-05-2012 & 00014044 & 5.02 & 1.37\\
J1354.2$-$3746  & 2MASXJ13541542$-$3746333 & 208.5642 & -37.7759 & Sy2 & 0.017  & 27-05-2012 & 00014045 & 5.02 & 0.42\\
J1419.2+0755  & 2MASXJ14190832+0754499 & 214.7846 & 7.9138 & Sy1 & 0.056  & 20-03-2012 & 00014046 & 5.02  &1.18\\
J1857.1$-$7829  & 2MASXJ18570768$-$7828212 & 284.2823 & -78.4726 & Sy1 & 0.042  & 16-12-2011 & 00014049 & 4.99 & 1.35\\
J2021.8+4400  & 2MASXJ20214907+4400399 & 305.4544 & 44.0110 & Sy2 & 0.017  & 21-04-2014 & 00016158 & 4.91 & 1.04\\
\hline
\hline
\end{tabular}}
\caption{\normalsize Summary of the sample of sources analyzed in this work. The SWIFT name is the one reported in the Palermo 100-month BAT catalog (Segreto et al. in prep.). R.A. and Decl are the right ascension and declination, taken by the SIMBAD Astronomical Database, of the counterpart of the 4PBC source. The counterpart name is reported in column 2. The source redshift and optical classification reported in columns 5--6 have been obtained from the SIMBAD Astronomical Database. The type classification is reported as follows: Sy1 - Seyfert 1 galaxy, Sy2 - Seyfert 2 galaxy, G - galaxy. Sources classified as galaxies in the \textit{Swift}-BAT survey are likely AGN for which an optical spectrum has not been acquired yet \citep{ajello12}. In columns 7--9 we report the date, ID and exposure time of the \cha\ ACIS-I observations studied in this paper. Finally, in column 10 we report the offset between the \cha\ source position and the BAT one.}
\end{table*}\label{tab:sample}

\section{Data Analysis}\label{sec:data_red}

On-board the \textit{Swift} satellite \citep{gehrels04} is the wide-field (120$\times$90 deg$^2$) Burst Alert Telescope \citep[BAT;][]{barthelmy05}. Since its launch, BAT has continuously observed the entire sky, covering the 15-150 keV energy range. In the most recent update to its survey (the BAT 100-month survey catalog, Segreto et al. in prep.), sources have been detected down to a flux limit $f\sim$3.3 $\times$ 10$^{-12}$ \flu in the 15-150 keV band\footnote{http://bat.ifc.inaf.it/100m\_bat\_catalog/100m\_bat\_catalog\_v0.0.htm}. Considering BAT high sensitivity and full-sky coverage, it provides an excellent view of the low-redshift, hard X-ray low-luminosity population.

The BAT\_IMAGER code \citep{segreto10} was used to reprocess the BAT survey data available in the HEASARC public archive and generate the 15-100\,keV spectra used in this work,
that were obtained by averaging over the whole BAT exposure. We utilized the official BAT spectral redistribution matrix\footnote{http://heasarc.gsfc.nasa.gov/docs/heasarc/caldb/data/swift/\\bat/index.html}.

In Table \ref{tab:sample} we report the list of 14 sources we analyze in this work.
As can be seen, most of the target sources (8 out of 14) are at redshift lower than 0.04, where the vast majority ($\sim$85\%) of the BAT-detected CT-AGN has been detected so far  \citep[see, e.g.,][]{burlon11,ricci15}.

It is worth noticing that all fourteen objects have available archival XRT observations with different exposure times. Five of these exposures (i.e., those of NGC 1125, ESO 424$-$12, 2MASXJ06403799$-$4321211, 2MASXJ18570768$-$7828212, and 2MASXJ20214907+4400399) were taken, albeit off axis, before the {\it Chandra} observations. In this work we use three of these observations to study potential long-term variability in three sources (see Section \ref{sec:nustar}). For the remaining eleven objects, however, we find that only four sources have XRT observations longer than 5\,ks, none of which provides sufficient count statistics to improve our fits.

\subsection{Chandra Source Analysis}

Level 2 event data was retrieved from the \textit{Chandra} Data 
Archive\footnote{\href{http://cda.harvard.edu/chaser}{http://cda.harvard.edu/chaser}} and reduced with the CIAO
\citep{fruscione06} 4.6 software and the \textit{Chandra}
Calibration Database (\textsc{caldb}) 4.5.1.1, adopting standard procedures.
After excluding time intervals of background flares exceeding \(3\sigma\) with 
the \textsc{lc\_sigma\_clip} task, we obtained the low-background total exposures 
listed in Table \ref{tab:sample}. 

Source identification in \textit{Chandra} images was unambiguous, with the source associated in the 
100-month Palermo BAT Catalogue being the brighter (if not the only) source in the \textit{Chandra} field.
 The only exception is J1354.2$-$3746, that has no associated optical/infrared counterpart in the 100-month Palermo BAT Catalog. 
Within the field of view of \cha\ observation 00014045, centered on J1354.2$-$3746 \swi\ position, there are six \cha\ sources: one of them is associated to the infrared source 2MASX J13541542 3746333, while the remaining five are associated to faint SDSS objects. While these latter do not exceed $\sim$60 net counts in the 0.3-8 keV range, 2MASX J13541542-3746333 show more than 600 net counts in the same Chandra band. Consequently, we chose 2MASX J13541542-3746333 and the associated \cha\ source as the J1354.2$-$3746 \swi\ counterpart.

The sources show no significant pile up, as measured by the CIAO \textsc{pileup\_map} tool.
\textit{Chandra}-ACIS spectra were extracted using the CIAO \textsc{specextract} task.
Source spectra were extracted in circular regions of 5$^{\prime\prime}$ radius centered at the
source coordinates, while background extraction has been performed in annuli with inner radius and outer
radii of 5$^{\prime\prime}$ and 15$^{\prime\prime}$, respectively. For the source spectra we applied the point-source aperture correction to the
\textsc{specextract} task. To make use of the \(\chi^2\) fit 
statistic we binned the spectra to obtain a minimum of \(20\) counts per bin. 

\subsection{\nustar\ Data Reduction}
For five out of fourteen sources in our sample there are available \nustar\ observations. These sources are ESO 33$-$2, 2MASXJ06403799$-$4321211, IC 2461, Was 49 and 2MASXJ18570768$-$7828212. The details of the \nustar\ observations for these objects are listed in Table \ref{tab:nustarobs}.

The data retrieved for both  \nustar\  Focal Plane Modules \citep[FPMA and FPMB;][]{harrison13} were processed using the  \nustar\  Data Analysis Software (NUSTARDAS) v1.5.1. 
The event data files were calibrated running the {\tt nupipeline} task using the response file from the Calibration Database (CALDB) v. 20100101. With the {\tt nuproducts} script we generated both the source and background spectra, and the ancillary and response matrix files. 
For both focal planes, we selected the source with a circular extraction region of diameter of 30$^{\prime\prime}$ centered on the target source; for the background we used the same extraction region positioned far from any source contamination in the same frame.

\begin{table*}[t!]
\begin{center}
\resizebox{\textwidth}{!}{ 
 \begin{tabular}{llccc}
 \hline
  \hline
 SWIFT Name & Source Name & Obs. Date & Obs. ID & Exposure time \\
 4PBC & & D-M-Y & & ks \\ 
 (1) & (2) & (3) & (4) & (5) \\
 \hline
 J0455.8$-$7531 & ESO 33-2 & 04--05--2014 & 60061054002 & 23.6\\
 J0640.6$-$4320 & 2MASXJ06403799$-$4321211 & 21--01--2014 & 60061070002  & 22.0 \\
 J0920.0$+$3711& IC 2461 & 13--06-2014 & 60061353002 & 32.9\\
 J1214.2+2932 & Was 49 & 15--07--2014 & 60061335002 & 20.4\\
 J1857.1$-$7829 & 2MASXJ18570768$-$7828212 & 27--07--2013 & 60061290002 & 18.0\\
 \hline
  \hline
\end{tabular}}
\end{center} \caption{\normalsize Summary of the \nustar\ observations available for five out of fourteen sources in our sample. The SWIFT name is the one reported in the Palermo 100-month BAT catalog (Segreto et al. in prep.), from which we get also the counterpart name. The \nustar\ observation date, observation ID and exposure are reported in columns (3), (4) and (5), respectively.}\label{tab:nustarobs} 
\end{table*}

\section{Spectral Analysis}\label{sec:analysis}

In this section we present the analysis of the X-ray spectra of the 14 sources in our sample. We detail the process of building an appropriate model, the characteristics of the models used to fit our composite spectral data, and elaborate on the implied physical significance of each model. For the spectral analysis we utilized XSPEC v.12.8.2 \citep{arnaud96}. Galactic absorption along the line of sight was determined for each source \citep{kalberla05}. 

In Table \ref{tab:best-fit} we report a summary of the best-fit models we obtained for each source in our sample. To determine the best-fit model for each source, we used an iterative process: we started from a basic model, i.e., an absorbed powerlaw, and progressively added further components. To verify the significance of an additional component, we performed a F-test, if statistically allowed. When it was not possible to statistically assess the improvement of the fit with a F-test \citep[i.e., when adding a Gaussian or a reflection component, see ][]{protassov02}, we kept the additional component if the reduced $\chi^2$ ($\chi^2_\nu$=$\chi^2$/degrees of freedom) significantly improved. In all four fits with either a Gaussian or a reflection component, the reduced $\chi^2$ without the additional component was $\chi^2_\nu>$1.5 and decreased to $\chi^2_\nu\sim$1--1.2.
 
For our spectral analysis we made use of the following components. 
\begin{enumerate}
\item All sources were fitted with a basic power-law to model the intrinsic nuclear AGN emission. In all but one source (Mrk 1179), this power-law was absorbed, i.e., we measured a significant value of intrinsic absorption ($N_{\rm H}$) caused by the gas surrounding the AGN accretion disk and the hot corona responsible for the X-ray emission.
\item Nine out of fourteen spectra (IC 1816, NGC 1125, ESO 33$-$2, ESO 424$-$12, Was 49, 2MASXJ13541542$-$3746333, 2MASXJ14190832+0754499, 2MASXJ18570768$-$7828212 and 2MASXJ20214907+4400399) required a second power law, having the same photon index of the first. This second power-law describes the emission component scattered, rather than absorbed by the material surrounding the accreting supermassive black hole \citep[SMBH; see, e.g.,][]{winter09}; a further discussion of this type of model is reported in Section \ref{sec:2pl}.
\item One source in the sample (Was 49) showed evidence of an excess in the spectrum at 20--40\,keV, thus indicating the presence of a reflection hump. This source was fitted with the \texttt{pexrav} \citep{magdziarz95} model, which we extensively describe in Section \ref{sec:refl}.
\item Three fits (those of IC 1816, ESO 33$-$2 and IC 2461) significantly improved adding a narrow ($\sigma$=50\,eV) Gaussian to model an excess in the spectrum at energy $E\sim$6.4\,keV (rest-frame), likely due to the Iron K$\alpha$ line. We report the best-fit values for the Iron K$\alpha$ equivalent width (EW), computed using the \texttt{eqwidth} tool in XSPEC, in Table \ref{tab:fit_results}.
\item One source (IC 1816) showed a spectral bump at energies below 1\,keV. IC 1816 is a heavily obscured AGN and it is therefore unlikely that this excess is caused by warm gas surrounding the AGN emission disk or the broad line region, as commonly observed in unobscured AGN \citep[see, e.g.,][]{risaliti04}. However, IC 1816 has been reported to be a starburst galaxy \citep{schmitt99,gu01}, and the source measured observed luminosity in the 0.3--1\,keV band (log(L$_{\rm 0.3-1keV}$)$\sim$40) is consistent with being due to star-forming emission processes. We modelled this soft excess with a phenomenological thermal component, using the XSPEC model \texttt{mekal} \citep[e.g.,][]{mewe85}. We find a best-fit temperature kT$<$0.8 keV.

\item For each of our sources, we utilize a composite spectra from the observations of two or three different observatories, which were not taken simultaneously. In acknowledgement of potential long-term flux variability in our objects, we used a multiplicative constant in each of our models. The constant was held to 1 for the BAT spectra and allowed to vary for the \textit{Chandra} and \textit{NuSTAR} spectra. For all the five sources with \nustar\ data, leaving the \nustar\ constant free to vary does not significantly improve the fit. Therefore, in all these five objects we fixed $K_{\rm Nus}$ to 1.
For a minority of objects, leaving the \cha\ constant free to vary also does not significantly improve the fit: for these sources, we therefore keep the constant frozen to $K_{\rm Cha}$=1. In Table \ref{tab:fit_results} we report the constant best-fit value for all those sources where the fit was significantly improved by leaving the constant free to vary. We further discuss this variability in Section \ref{sec:nustar}.

\end{enumerate}

\begin{figure}
\centering
\includegraphics[width=0.5\textwidth]{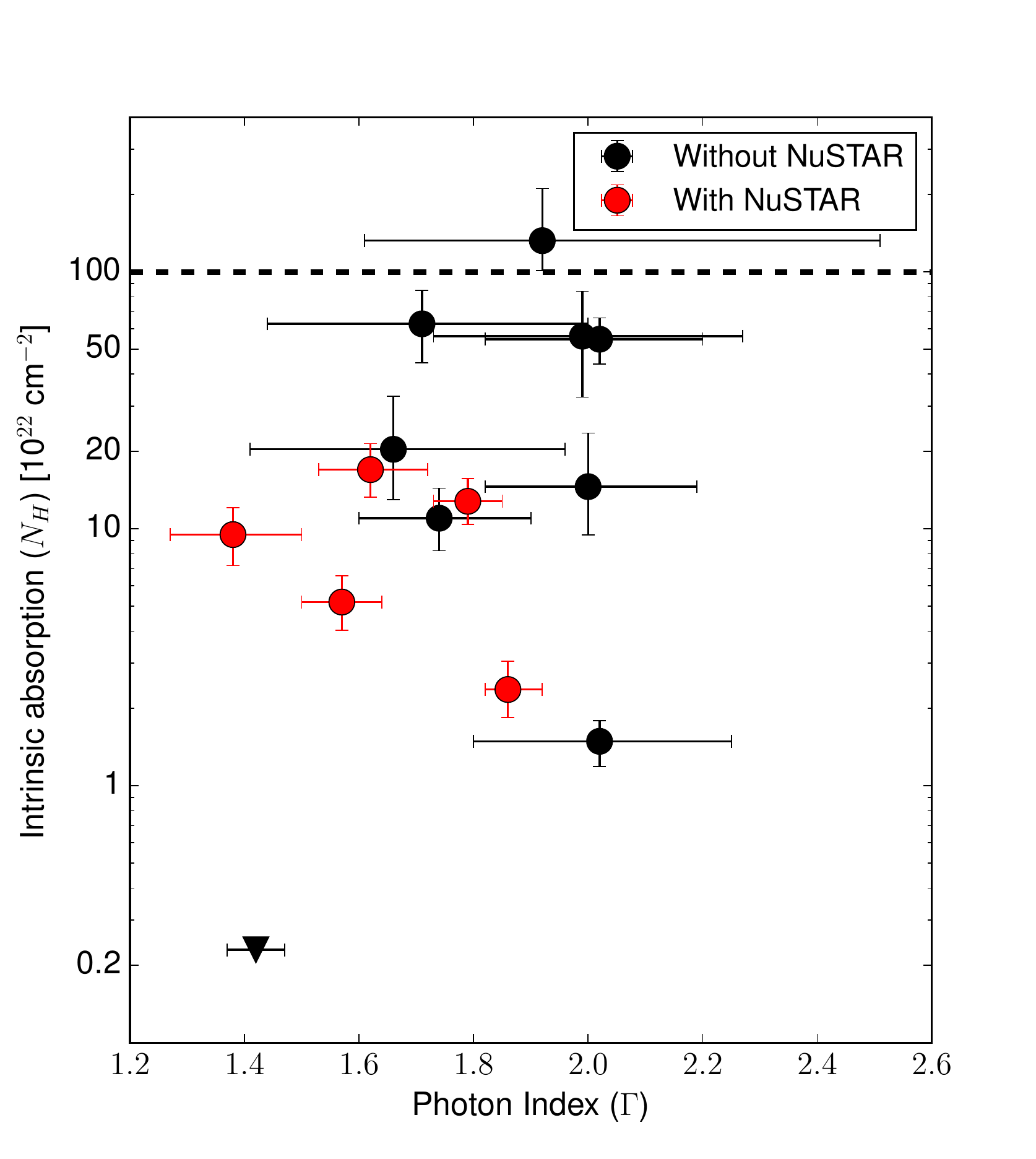}
\caption{\normalsize Intrinsic absorption $N_{\rm H}$ as a function of photon index for the 14 sources in our sample. Values for sources with \nustar\ spectra are plotted in red, while sources with only \cha\ and \swi\ data available are plotted in black. Upper limits on $N_{\rm H}$ are plotted as downwards triangles. The dashed horizontal line marks the $N_{\rm H}$=10$^{24}$ cm$^{-2}$ value, above which sources are classified as Compton thick AGN.}
\label{fig:gamma_nh}
\end{figure}

The best-fit parameters for each object in our sample are reported in Table \ref{tab:fit_results}, while in Figure \ref{fig:gamma_nh} we report the $N_{\rm H}$ best-fit values as a function of the $\Gamma$ ones. As can be seen, the majority of sources have moderate to high levels of intrinsic absorption (nine out of fourteen sources have $N_{\rm H}>$10$^{23}$ cm$^{-2}$).

Interestingly, two sources (Mrk 1179 and 2MASSXJ06403799$-$4321211) show relatively hard photon indexes, having both $\Gamma\sim$1.4 instead of the typical AGN photon index $\Gamma$=1.7--1.8 \citep[see, e.g.,][]{marchesi16c}. This may be an indication that for these sources the $N_{\rm H}$ values are slightly underestimated and/or that the characterization of the $<$2\,keV spectra requires additional components, such as a thermal one, or multiple emission lines. However, the available data do not allow us to statistically assess the significance of these additional components, which would require longer 0.5--10\,keV observations to be measured.

\begingroup
\renewcommand*{\arraystretch}{1.2}
\begin{table*}
\centering
\scalebox{0.85}{
\begin{tabular}{lcl}
\hline
\hline
Source    &XSPEC Model & Interpretation \\
\hline
IC 1816  & MEKAL+POW$_2$+ZWABS*(POW+ZGA) & Significant absorption and evidence of Fe-K$\alpha$ line. Thermal component at $<$1\,keV.\\
Mrk 1179  & ZWABS*POW & No significant intrinsic absorption. \\
NGC 1125 & POW$_2$+ZWABS*POW & Absorbed direct emission component, unabsorbed scattered emission component.\\
ESO 33$-$2 & POW$_2$+ZWABS*(POW+ZGA) & Significant absorption and evidence of Fe-K$\alpha$ line.\\
ESO 424$-$12 & POW$_2$+ZWABS*POW & Intrinsically absorbed main component, unabsorbed scattered component. \\ 
2MASXJ06403799$-$4321211  & ZWABS*POW & Significant intrinsic absorption from dusty torus. \\
IC 2461  & ZWABS(POW + ZGA) & Intrinsic absorption on direct emission, Fe-K$\alpha$ line. \\
Was 49  & POW$_2$+ZWABS*PEXRAV & Reflected emission with intrinsic absorption, unabsorbed scattered component. \\
2MASSJ13393397+5546142  & ZWABS*POW & Significant intrinsic absorption from dusty torus. \\ 
2MASXJ13532820$-$1123055  & ZWABS*POW & Significant intrinsic absorption from dusty torus. \\
2MASXJ13541542$-$3746333  & ZWABS*POW$_2$+ZWABS*POW &  Both direct and scattered emission components affected by intrinsic absorption. \\ 
2MASXJ14190832+0754499  & ZWABS*POW$_2$+ZWABS*POW & Both direct and scattered emission components affected by intrinsic absorption.  \\
2MASXJ18570768$-$7828212  & ZWABS*POW$_2$+ZWABS*POW & Both direct and scattered emission components affected by intrinsic absorption.\\
2MASXJ20214907+4400399  & POW$_2$+ZWABS*POW & Absorbed direct emission component, unabsorbed scattered emission component.\\
\hline
\hline
\end{tabular}}
\caption{\normalsize Best-Fit Models and Physical Interpretations. We report a description of each model component in the text. Each model was fitted with an absorption component to account for the absorption of our own galaxy. The different component used in the models are extensively described in Sections \ref{sec:spl}, \ref{sec:2pl}, \ref{sec:refl} and \ref{sec:nustar}.}\label{tab:best-fit}
\end{table*}
\endgroup

\begingroup
\renewcommand*{\arraystretch}{1.5}
\begin{sidewaystable*}[]
\centering
\scalebox{0.87}{
\begin{tabular}{lcccccccccccc}
\hline
\hline
Source & $N_{H(Gal)}$ & $N_{H(abs)}$ & $\Gamma$ & $K$ & $R_{\rm scatt}$ & EW &  $R_{\rm refl}$ & Log$f_{2-10keV}$ & Log$f_{15-55keV}$ & Log$L_{2-10keV}$ & Log$L_{15-55keV}$ &  $\chi^2/DOF$ \\  %
& $10^{20}$cm$^{-2}$ & $10^{22}$\,cm$^{-2}$ & & & & keV & & erg s$^{-1}$ cm$^{-2}$ & erg s$^{-1}$ cm$^{-2}$ & erg s$^{-1}$ & erg s$^{-1}$\\
\hline
IC 1816  & 2.58 & 62.84$^{+21.91}_{-18.63}$ & 1.71$^{+0.29}_{-0.27}$ & \nodata & \nodata & 0.49$^{+0.52}_{-0.46}$ & \nodata & -12.09$^{+0.04}_{-0.50}$ & -11.10$\pm$0.03 & 41.72 & 42.72 & 12.9/15 \\ 
Mrk 1179  & 7.44 & $\leq$0.24 & 1.42$^{+0.05}_{-0.05}$ & \nodata & \nodata & \nodata & \nodata & -11.47$^{+0.02}_{-0.03}$  & -11.11$\pm$0.05 & 43.05 & 43.41 &155.8/148 \\ 
NGC 1125  & 2.70 & 132.27$^{+78.70}_{-31.27}$ & 1.92$^{+0.59}_{-0.31}$ & 1.68$^{+15.25}_{-1.46}$ & 0.007$^{+0.003}_{-0.003}$ & \nodata & \nodata & -12.18$^{+0.08}_{-0.95}$ & -11.16$\pm$0.04 & 41.62  & 42.28 &13.6/11 \\
ESO 33$-$2$^{N}$  & 8.90 & 2.37$^{+0.69}_{-0.53}$ & 1.86$^{+0.06}_{-0.04}$ & 0.32$^{+0.05}_{-0.04}$ & 0.007$^{+0.003}_{-0.004}$ & 0.13$^{+0.45}_{-0.08}$ & \nodata & -11.50$^{+0.04}_{-0.05}$ & -11.05$\pm$0.03 & 42.31  & 42.84 & 439.1/371\\
ESO 424$-$12  & 2.17 & 54.68$^{+11.47}_{-10.92}$ & 2.02$^{+0.18}_{-0.20}$ & \nodata &  0.007$^{+0.010}_{-0.004}$ & \nodata & \nodata & -11.63$^{+0.03}_{-0.18}$ & -11.11$\pm$0.02 & 42.43 & 42.23 & 19.2/25 \\
2MASXJ06403799$-$4321211$^{N}$  & 5.79 & 9.48$^{+2.66}_{-2.33}$ & 1.38$^{+0.12}_{-0.11}$ & 2.49$^{+0.93}_{-0.65}$ &  \nodata & \nodata & \nodata & -11.60$^{+0.08}_{-0.12}$ & -11.13$\pm$0.03 & 43.52 & 43.82 & 114.0/115 \\
IC 2461$^{N}$ & 1.08 & 5.19$^{+1.38}_{-1.16}$ & 1.57$^{+0.07}_{-0.07}$ & 0.47$^{+0.07}_{-0.07}$ & \nodata & 0.31$^{+0.15}_{-0.15}$ & \nodata & -11.66$^{+0.05}_{-0.07}$ & -11.01$\pm$0.03 & 41.51 & 42.09 & 213.52/212 \\ 
Was 49$^{N}$ & 1.77 & 16.97$^{+4.47}_{-3.70}$ & 1.62$^{+0.10}_{-0.09}$ & \nodata &  0.081$^{+0.029}_{-0.015}$  & $<$0.18 & 1.6$^{+2.3}_{-1.1}$ & -11.84$^{+0.04}_{-0.04}$ & -11.20$\pm$0.03 & 43.12 & 43.83 & 101.5/91 \\
2MASSJ13393397+5546142 & 0.85 & 1.49$^{+0.30}_{-0.27}$ & 2.02$^{+0.23}_{-0.22}$ & 0.26$^{+0.17}_{-0.10}$ & \nodata & \nodata & \nodata & -11.68$^{+0.05}_{-0.78}$ & -11.34$\pm$0.05 & 43.91 & 44.26 & 70.1/72 \\ 
2MASXJ13532820$-$1123055  & 3.78 & 20.42$^{+12.46}_{-7.38}$ & 1.66$^{+0.30}_{-0.25}$ & 0.97$^{+2.32}_{-0.64}$ & \nodata & \nodata &\nodata & -11.79$^{+0.06}_{-0.83}$ & -11.22$\pm$0.05 &  43.23 & 43.84 & 13.4/12 \\ 
2MASXJ13541542$-$3746333  & 6.11 & 1.96$^{+0.65}_{-0.65}$/14.55$^{+9.00}_{-5.13}$ & 2.00$^{+0.19}_{-0.18}$ & 0.58$^{+0.27}_{-0.18}$ & 0.347$^{+0.550}_{-0.243}$ & \nodata & \nodata & -11.35$^{+0.05}_{-1.50}$ & -11.09$\pm$0.04 &  42.45 & 42.73 & 96.05/83 \\ 
2MASXJ14190832+0754499 & 2.24 & 1.11$^{+0.35}_{-0.29}$ /56.20$^{+27.79}_{-23.63}$ & 1.99$^{+0.28}_{-0.26}$ & 0.41$^{+0.75}_{-0.22}$ &  0.206 $^{+0.773}_{-0.157}$ & \nodata & \nodata & -11.75$^{+0.05}_{-1.08}$ & -10.77$\pm$0.05 & 43.11 & 44.11 & 54.4/54 \\
2MASXJ18570768$-$7828212$^{N}$  & 8.33 & 1.27$^{+0.35}_{-0.33}$ /12.75$^{+2.92}_{-2.41}$ & 1.79$^{+0.06}_{-0.06}$ & \nodata & 0.268$^{+0.331}_{-0.156}$ & \nodata & \nodata & -11.27$^{+0.04}_{-0.73}$ & -10.96$\pm$0.02 &  43.33 & 43.66 & 402.9/294  \\
2MASXJ20214907+4400399  & 87.7 & 10.97$^{+3.39}_{-2.78}$ & 1.74$^{+0.16}_{-0.14}$ & \nodata & 0.027$^{+0.008}_{-0.006}$ & \nodata & \nodata & -11.54$^{+0.04}_{-0.28}$ & -11.25$\pm$0.06 & 42.26 & 42.56 & 53.58/47 \\ 
\hline
\hline
\end{tabular}}
\caption{\normalsize Summary of the best-fit values for the different parameters studied in this work. For sources with two intrinsic absorption ($N_{H(abs)}$) values, the higher value represents the absorption component of the emission that is absorbed by the dusty torus, while the second component affects the scattered emission component and is probably caused by the host galaxy. $K$ is a cross-normalization constant component $K$=$\frac{f_{Cha}}{f_{NuS,BAT}}$, added to the \cha\ spectrum to account for potential variability between the \cha\ and the \nustar\ and \swi\ observations. When the constant is not reported, leaving the constant free to vary did not improve the fit significantly. We find that adding a constant between the \nustar\ and the \swi\ data does not significantly improve the fit in any of the five objects with \nustar\ data. EW is the equivalent width of the Iron K$\alpha$ line. $R_{\rm scatt}$ is the ratio between the normalizations of the scattered component and of the main one. Sources flagged with $N$ have \nustar\ data available. The best-fit observed, not Galactic absorption-corrected fluxes ($f$), and unabsorbed luminosities ($L$) in the 2--10\,keV and 15--55\,keV bands are also reported.
}\label{tab:fit_results}
\end{sidewaystable*}
\endgroup

\subsection{Unabsorbed and absorbed Single Power-law Sources}\label{sec:spl}
Of the 14 sources in our sample, five (36\%) are best fitted with a single power-law (POW), with or without a significant intrinsic absorption component (ZWABS).
Only one source (Mrk 1179) showed no indication of significant intrinsic absorption, with a 90\% confidence level upper limits on $N_{\rm H}$ lower than 10$^{22}$ cm$^{-2}$.
The remaining objects revealed a statistically significant intrinsic absorption and were best fitted by the ZWABS*POW model. The intrinsic absorbing column density varies in the range $N_{\rm H}$=1--20 $\times$ 10$^{22}$ cm$^{-2}$. 

Finally, one of the sources fitted with a single power-law, i.e., IC 2461, also required an additional gaussian (GA) to model the excess at $\sim$6.4\,keV related to the Iron K$\alpha$ line, as discussed in the previous section.

\subsection{Double Power-law Sources}\label{sec:2pl}
Nine spectra (64\% of the sample) showed a significantly improved $\chi^2$ when a second power-law was added to the model. This power-law (POW$_2$) has the same photon index of the main power-law and accounts for a fraction of the main X-ray emission scattered, rather than absorbed, by the absorbing material surrounding the SMBH. The normalization of this second power-law is usually significantly smaller than the one of the main one, being typically between 1\,\% and 20\,\% \citep[see, e.g.][]{ueda14}. This is the case also for the majority of the sources in our sample (see in Table \ref{tab:fit_results} the ratio $R_{\rm scatt}$ between the secondary and main power-law normalization): for one object, 2MASXJ13541542$-$3746333, we find a slightly higher value of $R_{\rm scatt}$=0.347$^{+0.550}_{-0.243}$, but as can be seen the uncertainties are quite large and the ratio is consistent with $\sim$10\% within the 90\% confidence uncertainty.

Sources best-fitted by a double power-law, where a scattered component is observed, are usually significantly obscured \citep[see, e.g.,][]{marchesi16c}, and eight out of nine sources in this subsample have in fact N$_H>$10$^{23}$ cm$^{-2}$. We point out that (as can be seen in Table \ref{tab:fit_results}) three sources showed evidence of low, but significant obscuration ($N_H\sim$10$^{22}$ cm$^{-2}$) affecting the scattered component. This second absorbing component cannot be caused by the same obscuring material responsible for the absorption of the main power-law, since the ionized gas responsible for the scattering is thought to extend to size scales larger than those of the obscuring material \citep[see, e.g.,][]{turner97,ueda07}. Therefore, this obscuration can be linked to gas and dust in the AGN host galaxy.

\subsection{Sources with a significant reflected component}\label{sec:refl}
One source in our sample, Was 49, is best fitted using the \texttt{pexrav} model, developed by \citet{magdziarz95} to describe sources with reflected components \citep{winter09}. The only parameter of this model that we left free to vary, besides the photon index and the normalization, is the reflection intensity $R$, which is constrained to be $R\geq$0. The energy cut-off of the spectrum is fixed to E=300\,keV, the metal abundances are fixed to solar and the reflecting material has inclination angle $\theta$=60\degree.

Was 49 shows an excess in the \textit{Swift}-BAT and \nustar\ spectra, in the 20--50\,keV energy range. We find that $R$ is consistent with 1, i.e., with the case where the reflection is produced by an infinite slab isotropically illuminated by the corona emission.

\subsection{A candidate Compton-Thick AGN}\label{sec:ct}
\begin{figure}
\centering
\includegraphics[width=0.5\textwidth]{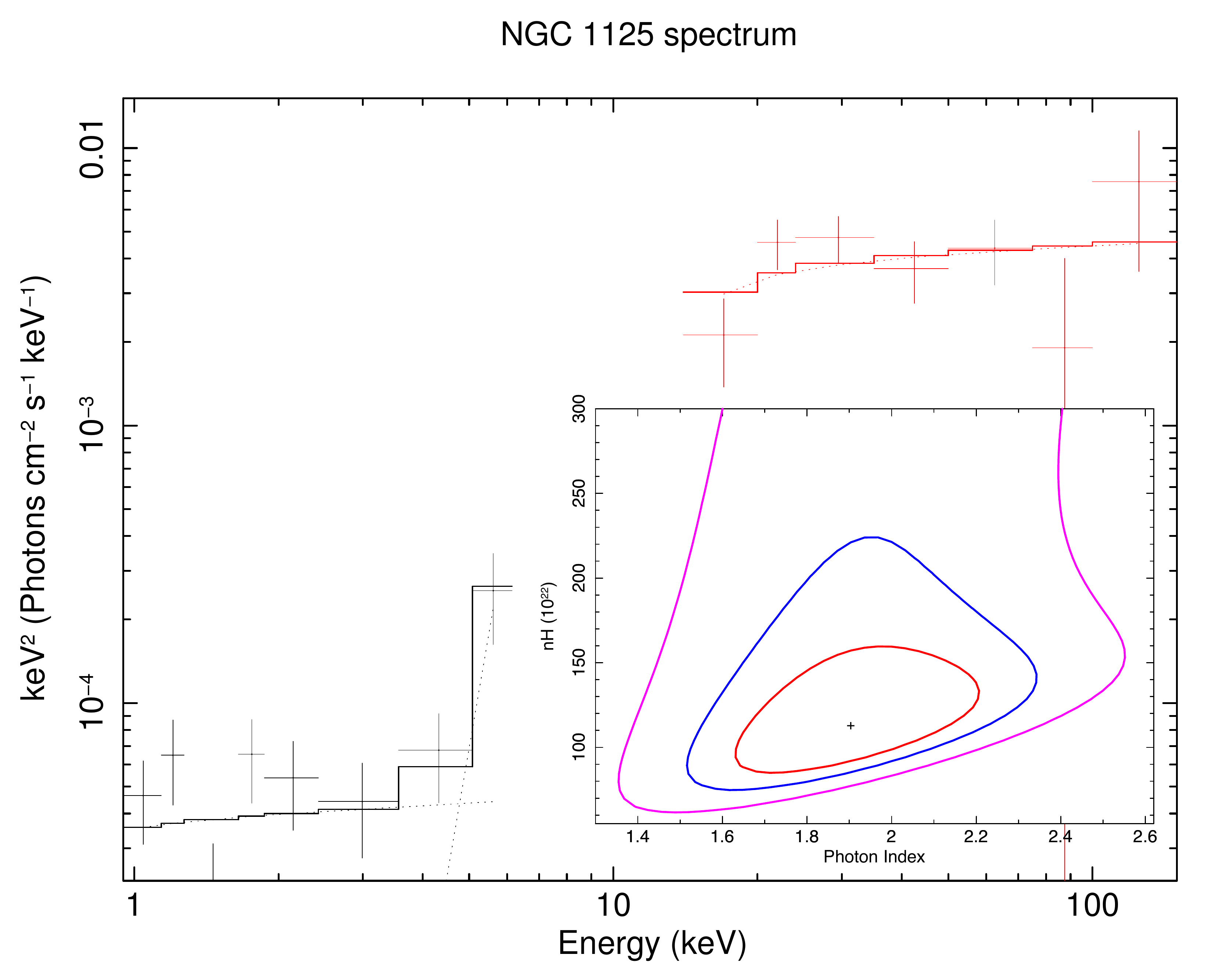}
\caption{\normalsize Unfolded spectrum from Chandra (black) and \textit{Swift}-BAT (red) of NGC 1125, the candidate CT-AGN in our sample. The confidence contours on the intrinsic absorption ($N_{\rm H}$) and photon index are shown in the inset.}
\label{fig:ct_spec}
\end{figure}

One of the objects in our sample, NGC 1125, is found to be a transmission dominated candidate CT-AGN, i.e., a CT source where a fraction of the intrinsic continuum pierces trough the obscuring material and we are able to measure $N_{\rm H}$ using the absorption turnover in the X-ray spectrum \citep[see, e.g.,][]{comastri11,georgantopoulos13}. NGC 1125 has best-fit intrinsic absorption N$_H$=1.32$^{+0.79}_{-0.31}\times$10$^{24}$ cm$^{-2}$. We report the spectrum of this object in Figure \ref{fig:ct_spec}; in the inset, the confidence contours on the photon index $\Gamma$ and on $N_{\rm H}$ are also plotted. As can be seen, the emission in the 0.5--7\,keV band is strongly depleted, and the flux in the 15--150\,keV band sampled by \swi\ is $\sim$40 times larger than the flux in the 0.5--7\,keV band observed with \cha.

It is worth noticing that CT-AGN are usually expected to have a prominent Iron K$\alpha$ feature at 6.4\,keV, while in this source the fit is not significantly improved by the addition of an emission line, and even obtaining a reasonable upper limit on the line EW is not possible. This is due to the fact that the source has a very low counts statistics ($\sim$30 counts in the 2--7\,keV band), consistently with being heavily obscured. Furthermore, the \cha\ effective area strongly declines at E$>$7\,keV, therefore making difficult to properly constrain the Iron K$\alpha$ line using \cha\ for sources in the local Universe ($z<$0.1).

Notably, the Compton thickness of NGC 1125 has been reported also by \citet{ricci15}, using the \swi\ data in combination with \textit{Swift}-XRT data. Their intrinsic absorption measurement is in good agreement with ours, being N$_{\rm H}$=1.83$^{+2.06}_{-0.76}\times$10$^{24}$ cm$^{-2}$.

Recently, \citet{koss16} developed a new technique to identify CT-AGN in AGN with low counts statistics using \swi\ or \nustar\ data: their method is based on the curvature of the AGN spectrum between 14 and 50 keV, parametrized as follows:
\begin{equation}
SC_{BAT}=\frac{-3.42A-0.82B+1.65C+3.58D}{Tot},\label{eq:sc_bat}
\end{equation} 

where $A$, $B$, $C$, $D$ and $Tot$ are the count rates measured with \swi\ in the 14-20\,keV, 20-24\,keV, 24-35\,keV, 35-50\,keV and 14-50\,keV, respectively. $SC_{BAT}$=0.4 is the CT-AGN selection threshold: seven of the nine sources with $SC_{BAT}>$0.4 in the \citet{koss16} sample have $N_{\rm H, z}>$ 10$^{24}$ cm$^{-2}$; the remaining two are significantly obscured ($N_{\rm H, z}>$ 5 $\times$ 10$^{23}$ cm$^{-2}$).

We tested the spectral curvature method for NGC 1125 and we found a value SC$_{BAT}$=0.44$\pm$0.15, slightly above the CT threshold, although the uncertainties on $SC_{BAT}$ are significant ($\sim$30\%).

\subsection{The strategic role of \nustar\ data in the spectral characterization of obscured AGN}\label{sec:nustar}
\begin{figure*}
\begin{minipage}[b]{.5\textwidth}
  \centering
  \includegraphics[width=1.02\textwidth]{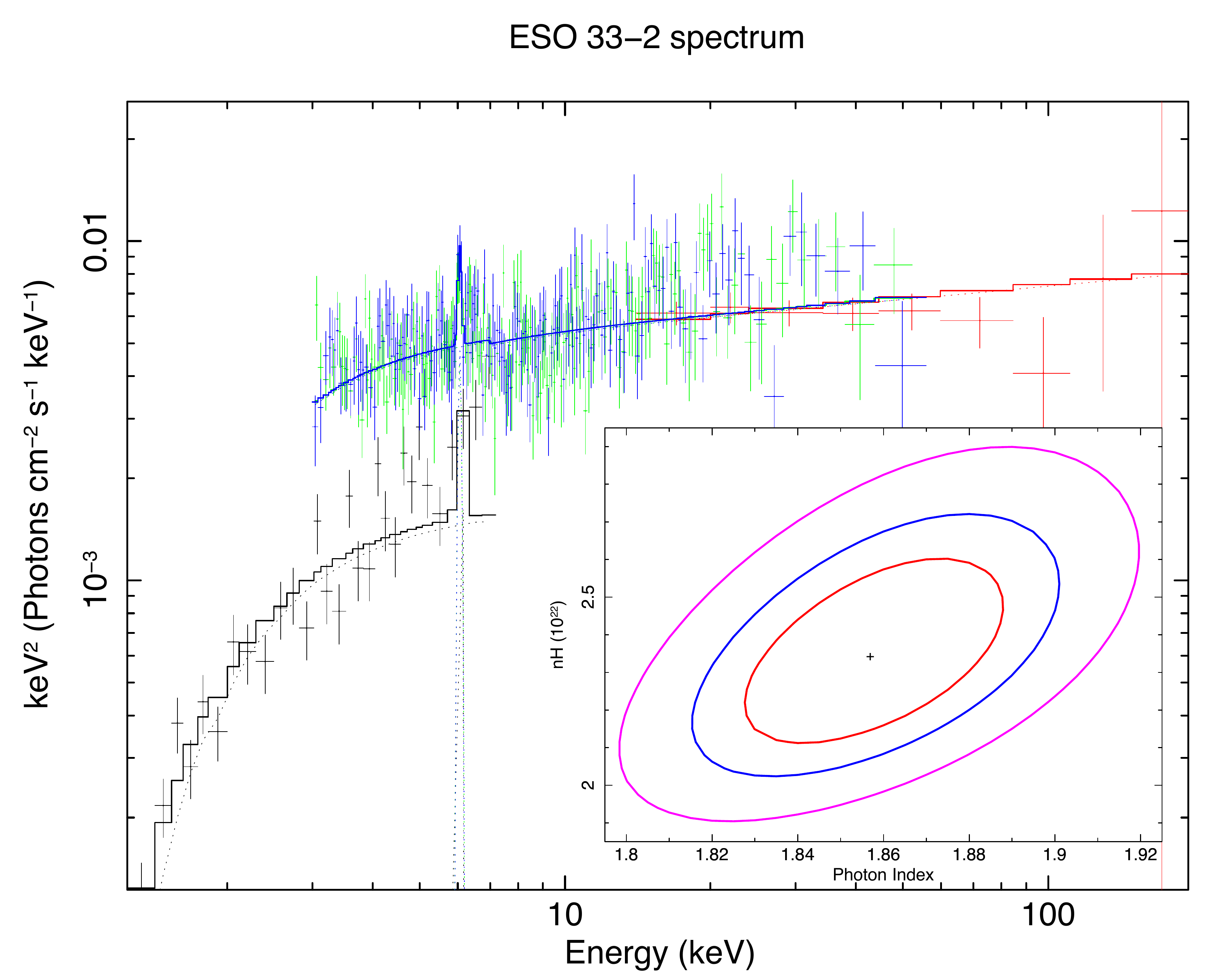}
  \end{minipage}
\begin{minipage}[b]{.5\textwidth}
  \centering
  \includegraphics[width=1.02\textwidth]{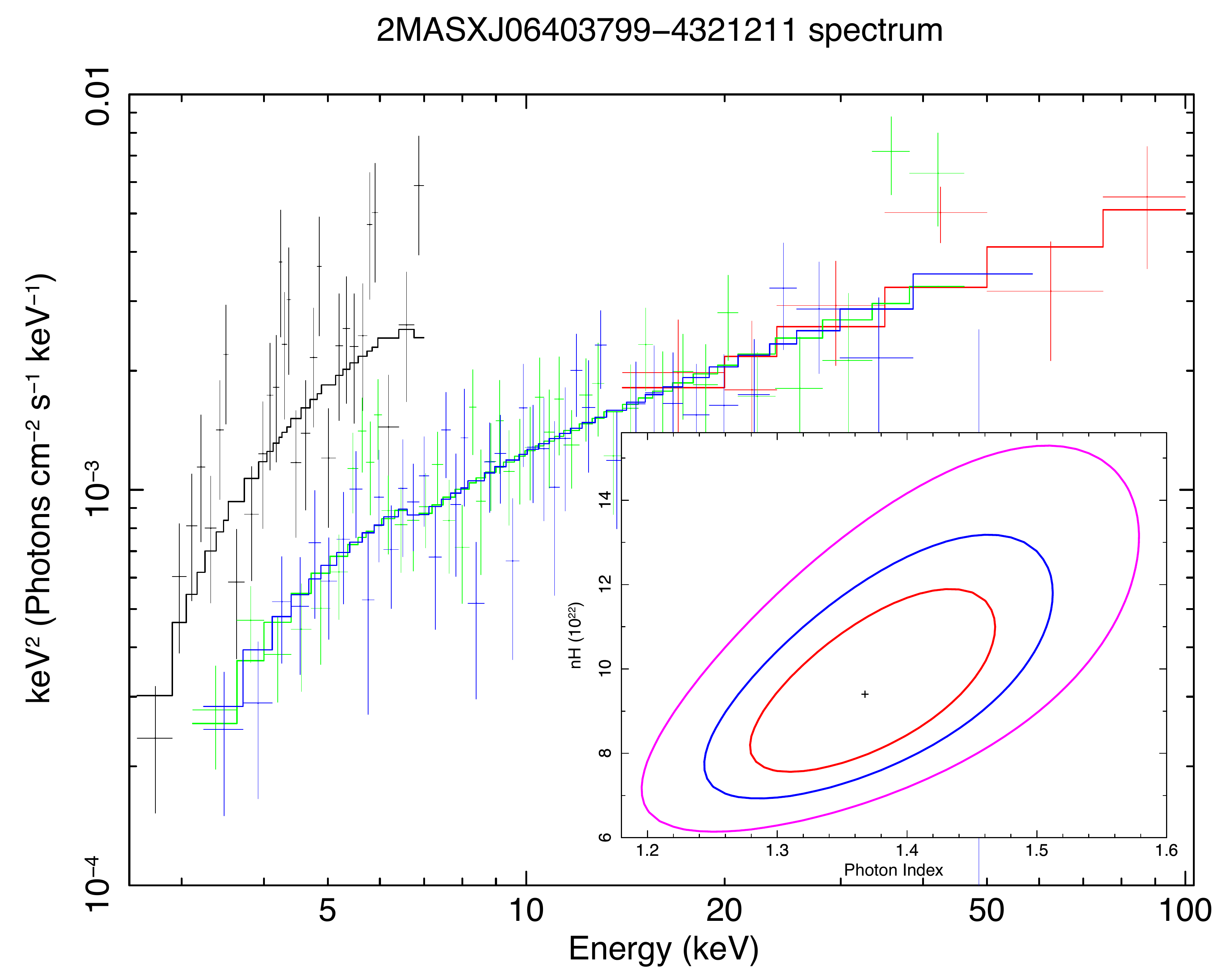}
  \end{minipage}
\begin{minipage}[b]{.5\textwidth}
  \centering
  \includegraphics[width=1.02\textwidth]{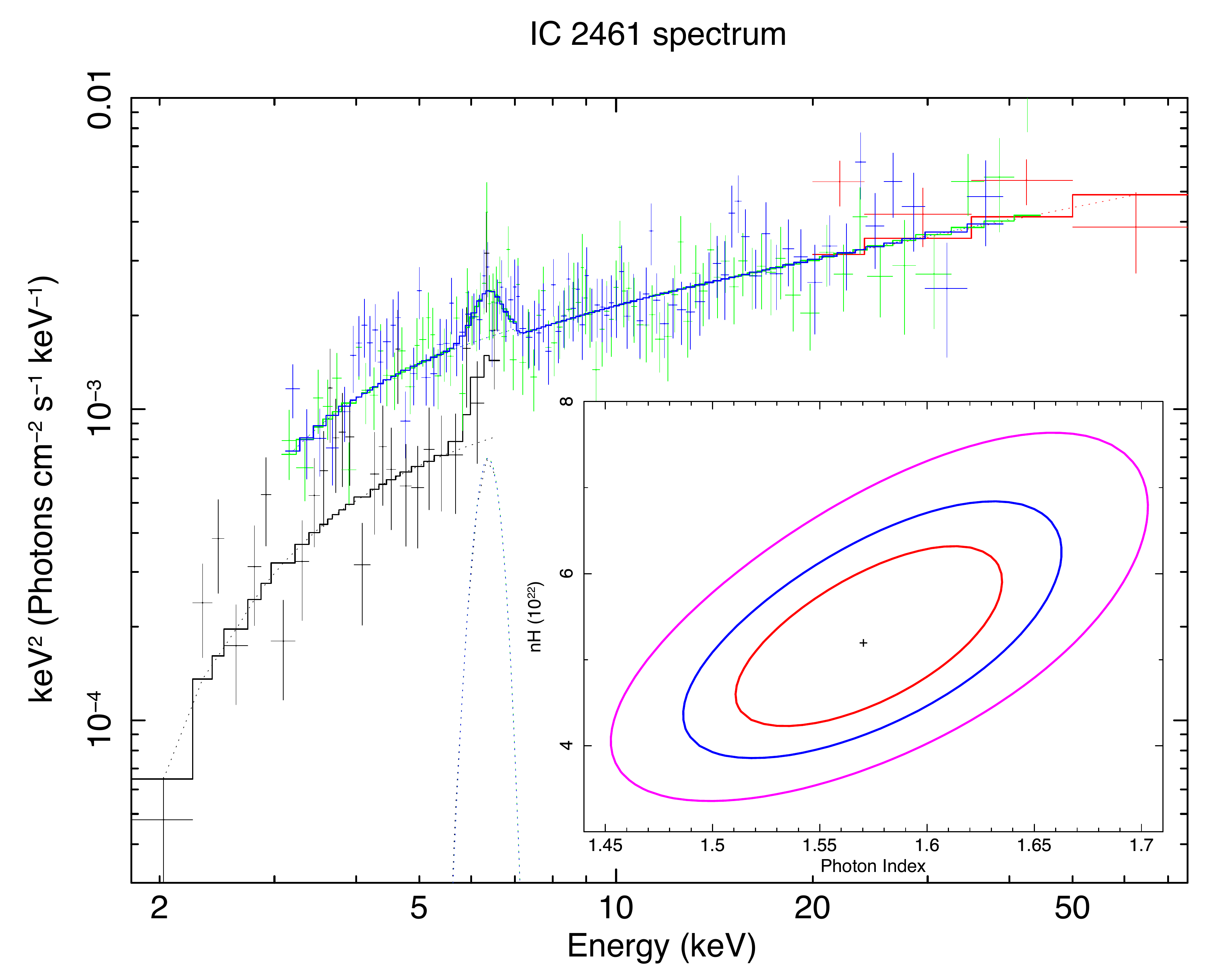}
  \end{minipage}
\begin{minipage}[b]{.5\textwidth}
  \centering
  \includegraphics[width=1.02\textwidth]{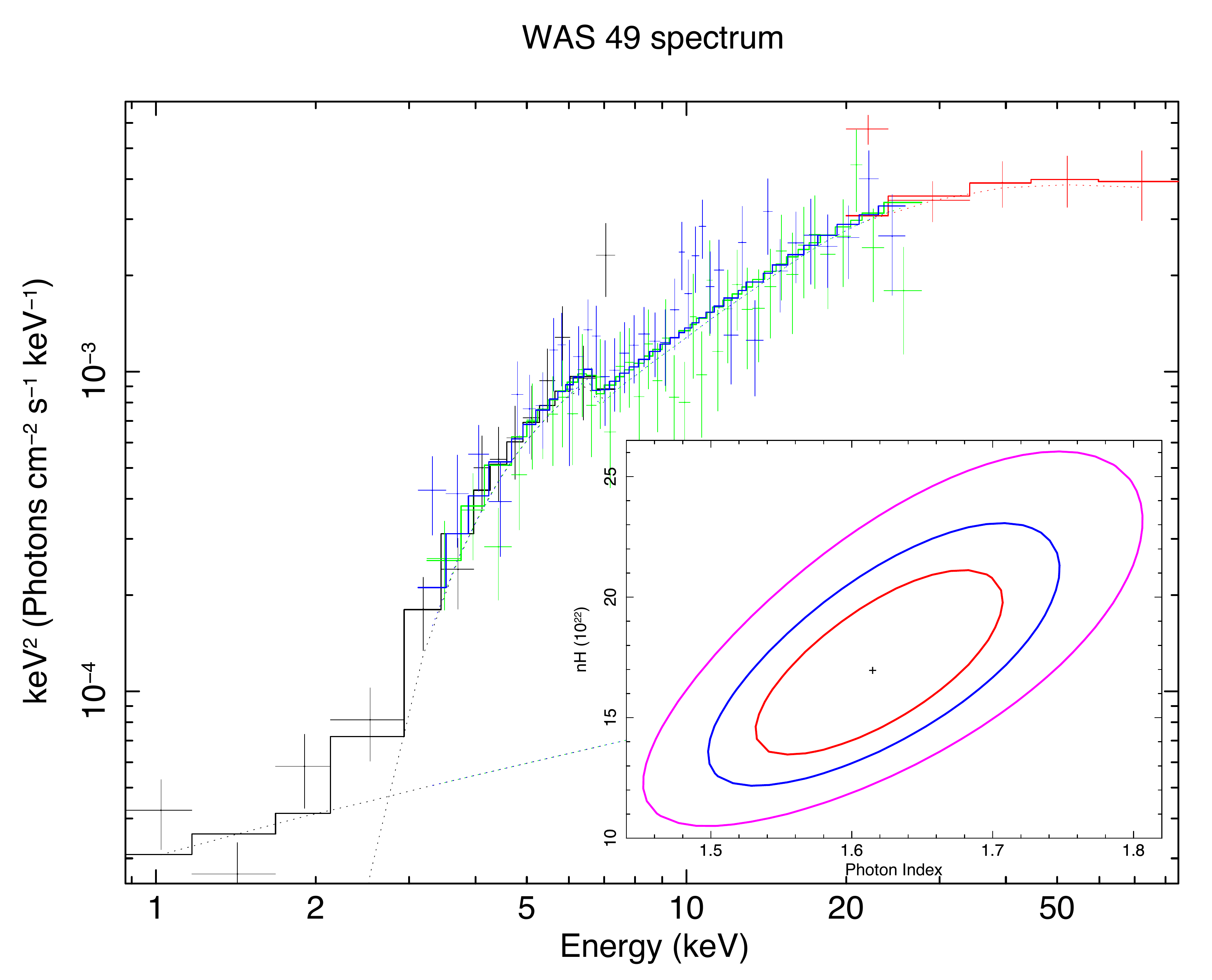}
  \end{minipage}
\begin{minipage}[b]{.5\textwidth}
  \centering
  \includegraphics[width=1.02\textwidth]{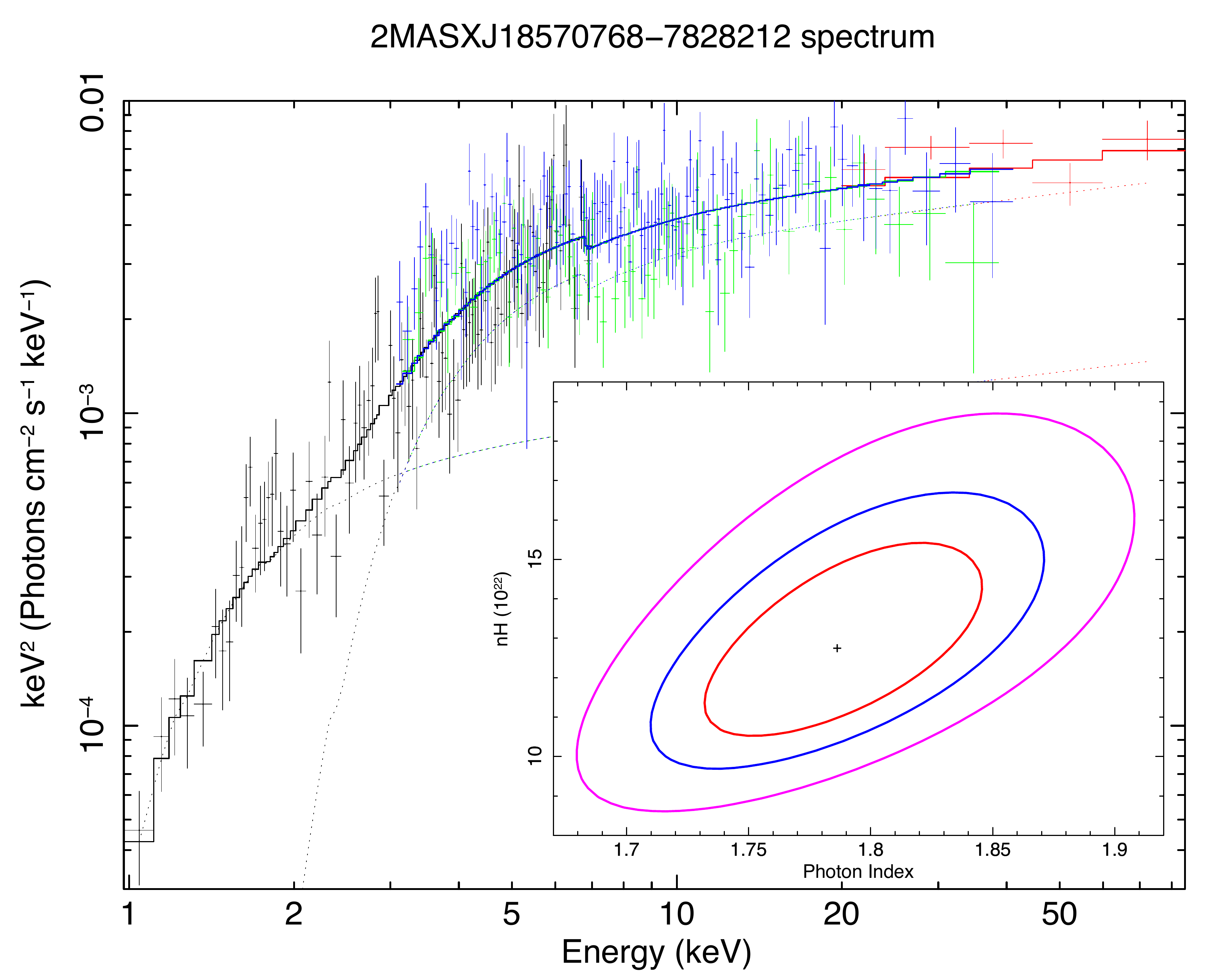}
  \end{minipage}
\caption{\normalsize \cha\ (black), \nustar\ (green and blue) and \swi\ (red) spectra of the five sources in our sample having an available \nustar\ observation. The best-fitting model is plotted as a solid line, while the single components are plotted as dotted lines. In the inset, the confidence contours at 68, 90 and 99\% confidence level for $\Gamma$ and $N_{\rm H}$ of the main emission component (in units of 10$^{22}$ cm$^{-2}$) are also shown.}\label{fig:nustar}
\end{figure*}

In Figure \ref{fig:nustar} we report the spectra of the five sources for which there are available \nustar\ observations. As can be seen, the \nustar\ data nicely cover the 3--40\,keV energy range, bridging the substantial gap between the \cha\ and \swi\ data. This offers the possibility to analyze with unprecedented quality the region of the spectrum where the reflection component is dominant.

At a face value, adding \nustar\ data to the \cha\ and \swi\ ones reduces the uncertainty on the photon index (from $\sim7$\,\% to $\sim$5\,\%) and on the intrinsic absorption N$_{\rm H}$ (from $\sim32$\,\% to $\sim$25\,\%).
\nustar\ data allows us to uncover the reflection component of Was 49, that is not constrained with \cha\ and \swi\ data alone. Was 49 \nustar\ spectrum has been analyzed also by \citet{secrest17}, finding best-fit photon index ($\Gamma$=1.6$\pm$0.1) and intrinsic absorption (N$_{\rm H}$=2.3$^{+0.5}_{-0.4}$$\times$10$^{23}$ cm$^{-2}$) in good agreement with the one we find.
For two objects, ESO 33$-$2 and IC 2461, \nustar\ allows us to uncover the presence of an Iron line K$\alpha$ line. In fact, to constrain the Iron line and where it originates from requires an accurate measurement of the continuum beneath the line, a task almost impossible using only short \cha\ observations, particularly at $\geq$7\,keV.

Among the nine objects in our sample with no available \nustar\ data, IC 1816 is the one where a \nustar\ follow-up will improve the most the characterization of the source. In fact, IC 1816 is a heavily obscured AGN ($N_{\rm H}$=6.28$^{+2.19}_{-1.86}\times$10$^{23}$ cm$^{-2}$) and the positive residuals observed on both sides of the strong Iron line can be interpreted as either the sign of a very broad Iron line (EW$\sim$4\, keV) or a mixture of the transmission and reflection components. It is not possible to determine the most likely scenario without a measurement of the continuum between \cha\ and \swi.

Three sources with \cha\ and \nustar\ data required the addition of a constant to the best-fit model, to take into account a significant difference in normalization between the \cha\ spectrum and the \nustar\ and \swi\ ones (see Figure \ref{fig:nustar}; as mentioned in Section \ref{sec:analysis}, we find that the addition of a constant to account for a normalization offset between the \nustar\ and \swi\ data does not produce a significant improvement of the fit in any of our sources). To verify if this difference is related to variability, i.e., an intrinsic change in the AGN flux between the \cha\ and \nustar\ observations, we used \textit{Swift}-XRT observations\footnote{Namely, \textit{Swift}-XRT ObsID 00080345001 for ESO 33$-$2, ObsID 00080376001 for 2MASXJ06403799$-$4321211 and ObsID 00080688002 for IC 2461} taken together with, or within 48 hours from, the \nustar\ observation. In all three cases we found that the \textit{Swift}-XRT flux is fully consistent with the \nustar\ one, therefore suggesting that the observed \textit{Chandra}--\nustar\ offset is most likely caused by AGN variability. In all cases, the variability between the epochs is mostly in flux of the intrinsic component (i.e., the variability is not likely due to variable absorption) and of the order of $\sim3$ at most, in agreement with typical variability amplitudes for radio-quiet AGN. 

We point out that one source, IC 2461, would have been incorrectly classified as CT-AGN according to the \cha\ and \swi\ data only. 
In fact, when fitting the IC 2461 spectrum without the \nustar\ data we find two possible solutions, which have similar statistics ($\chi^2_\nu\sim$1.2) and are both consistent with a CT scenario. In one case, the best-fit model is an absorbed power-law having CT intrinsic absorption $N_{\rm H, z}$=1.5$\pm$0.5$\times$10$^{24}$ cm$^{-2}$ and a relatively soft photon index, $\Gamma$=2.20$\pm$0.25. The second best-fit solution is instead obtained fitting the spectrum with \texttt{pexrav} \citep{magdziarz95}, for which we get a high reflection intensity value, $R>$10: such a strong reflection component is usually observed in heavily obscured AGN. However, both these scenarios are ruled out when the \nustar\ data are added to the fit.

Interestingly, IC 2461 would have erroneously been classified as a candidate CT-AGN using the spectral curvature equation for BAT (Equation \ref{eq:sc_bat}), having $SC_{BAT}$=0.49$\pm$0.11, above the $SC_{BAT}$=0.4 threshold. However, in the same work \citet{koss16} report a second parameterization of SC, based on the \nustar\ count rates in three different bands:

\begin{equation}
SC_{Nus}=\frac{-0.46 \times A+ 0.64 \times B+ 2.33 \times C}{Tot},
\end{equation} 

where $A$, $B$, $C$ and $Tot$ are the count rates measured with \nustar\ in the 8--14\,keV, 14--20\,keV, 20--30\,keV and 8--30\,keV bands, respectively. As for $SC_{BAT}$, $SC_{Nus}$=0.4 is the threshold adopted to select candidate CT sources. With this second parameterization, which is much more accurate than the BAT-based one, IC 2461 is correctly not classified as a CT-AGN, since its \nustar\ spectral curvature value is  $SC_{Nus}$=0.19$\pm$0.03.

Furthermore, for other two objects with \nustar\ data (2MASXJ06403799$-$4321211 and Was 49) the photon index $\Gamma$ value decreases by $\sim$15--25\% once the \nustar\ information is added, due to the significantly better statistics provided by \nustar\ in the 5--40\,keV band.

Finally, it is also interesting to note the good agreement between \nustar\ and \swi\ for all the cases tested.

\section{Discussion and Conclusions}\label{sec:discuss}
As discussed in the literature \citep{gilli07,treister09}, current cosmic X-ray background models require  20\% of the AGN population (at redshift z$\sim$0) to be Compton-thick. Because of the suppression of the AGN flux, detecting Compton-thick AGN has always been a challenging task.
The most efficient ways to discover heavily obscured AGN is to survey the sky at hard X-rays ($>$10\,keV), where part of the nuclear emission can pierce through the Compton-thick medium, or at infrared wavelengths where the intrinsic emission is reprocessed. To date only a handful of well studied, bona-fide, Compton-thick AGN exist \citep{comastri04,dellaceca08}.

Indeed, even above 10\,keV Compton down-scattering and absorption substantially degrade the intrinsic spectrum, making CT-AGN only a mere 5--10\,\% of the local AGN population \citep{ajello08b,burlon11,ricci15}. It is important to search in the \swi\ survey for heavily obscured AGN, despite the low turnout, because those are among the brightest AGN observable and once discovered they can be followed up, with relatively modest amounts of time, with focusing telescopes like \textit{Chandra}, XMM-\textit{Newton} and most importantly \nustar.

In this work we have presented the 0.3--150\,keV spectral analysis of 14 sources originally detected in the {\it Swift}-BAT 60\,month survey \citep{ajello12} that are also presented in the updated 100\,month BAT survey (Segreto et al. in prep.). These 14 sources were selected on the basis of their optical type (Sy2), the lack of a bright ROSAT counterpart or because they were the last few sources to be followed up of the 60\,month survey. As such there was an expectation that these sources may be absorbed. Indeed,all but one are absorbed AGN (N$_{\rm H}\geq 10^{22}$\,cm$^{-2}$, see Fig.~\ref{fig:gamma_nh}). 
Moreover, 9 out of 14 sources are heavily obscured ($N_H>$10$^{23}$ cm$^{-2}$), and one, NGC 1125, is a candidate CT-AGN ($N_H>$10$^{24}$ cm$^{-2}$). It is not surprising that the CT-AGN source is among the lowest redshift ones ($z$=0.011). Indeed, in CT-AGN the source flux is so heavily extinguished that \swi\ can only discover them at a redshift smaller than the average ($\langle z \rangle$=0.04 in our sample).

It is worth pointing out that our candidate CT-AGN is transmission dominated, i.e., some of its nuclear radiation pierces through the torus and the observed emission in the 0.5--7\,keV is almost entirely due to this scattered component. In our sample, we did not find any reflection dominated CT-AGN, i.e., a source where all the observed emission in the 0.5--7\,keV band is due to reflection, while the intrinsic emission is completely depleted by a dusty torus having column density N$_{\rm H}\geq 10^{25}$\,cm$^{-2}$.
Further studies have to be performed to constrain the ratio of transmission dominated on reflection dominated CT-AGN in the whole population of local CT-AGN, to properly characterize the typical geometry of the obscuring material surrounding nearby CT-AGN.

A key role can certainly be  played by \nustar\ in the proper characterization of these CT-AGN, especially of the reflection-dominated ones, where the Iron K$\alpha$ line is extremely prominent. As we show in this work, the \nustar\ data allows us to significantly reduce the uncertainties on quantities like $\Gamma$ and N$_{\rm H}$; more importantly, it put strong constraints on the presence and intensity of the Iron K$\alpha$ line at 6.4\,keV, since in this energy range the counts statistics for nearby AGN with low \cha\ (or XMM-\textit{Newton}) exposure is usually poor. \nustar\ spectra, instead cover with excellent statistics the energy range 3--20\,keV, therefore permitting to characterize CT-AGN with unprecedented spectral quality, studying parameters such as the Iron line, the continuum around the line itself and the strength of the reflection component.

Finally, in future studies, a further improvement should be made in the distance-based selection of source candidates in order to improve the fraction of heavily obscured AGN in the sample, since the majority of confirmed CT sources in the 100-month BAT catalog are at very low redshifts \citep{burlon11,ricci15}. Therefore, follow-ups to this study will likely be effective in detecting CT-AGN by selecting nearby, heavily obscurred AGN. Indeed, \citet{marchesi17} analyzed seven Seyfert 2 galaxies from the \swi\ survey, each lacked ROSAT counterparts in the 0.5--2.4\,keV band and where located at $z<$0.03. The result of this analysis was a new CT-AGN candidate and they determined that each of their seven sources are heavily obscured ($N_H>$10$^{23}$ cm$^{-2}$ at a 99\% confidence level). 

\section*{Acknowledgements}
We thank the anonymous referee, whose comments significantly improved the paper.

\begin{appendices}
\section{Fits and Residuals}
In this appendix we report all the spectra, best-fit models and associated data-to-model ratio for the fourteen sources in our sample.

\begin{figure*}
\begin{minipage}[b]{.5\textwidth}
  \centering
  \includegraphics[angle=270,width=1.02\textwidth]{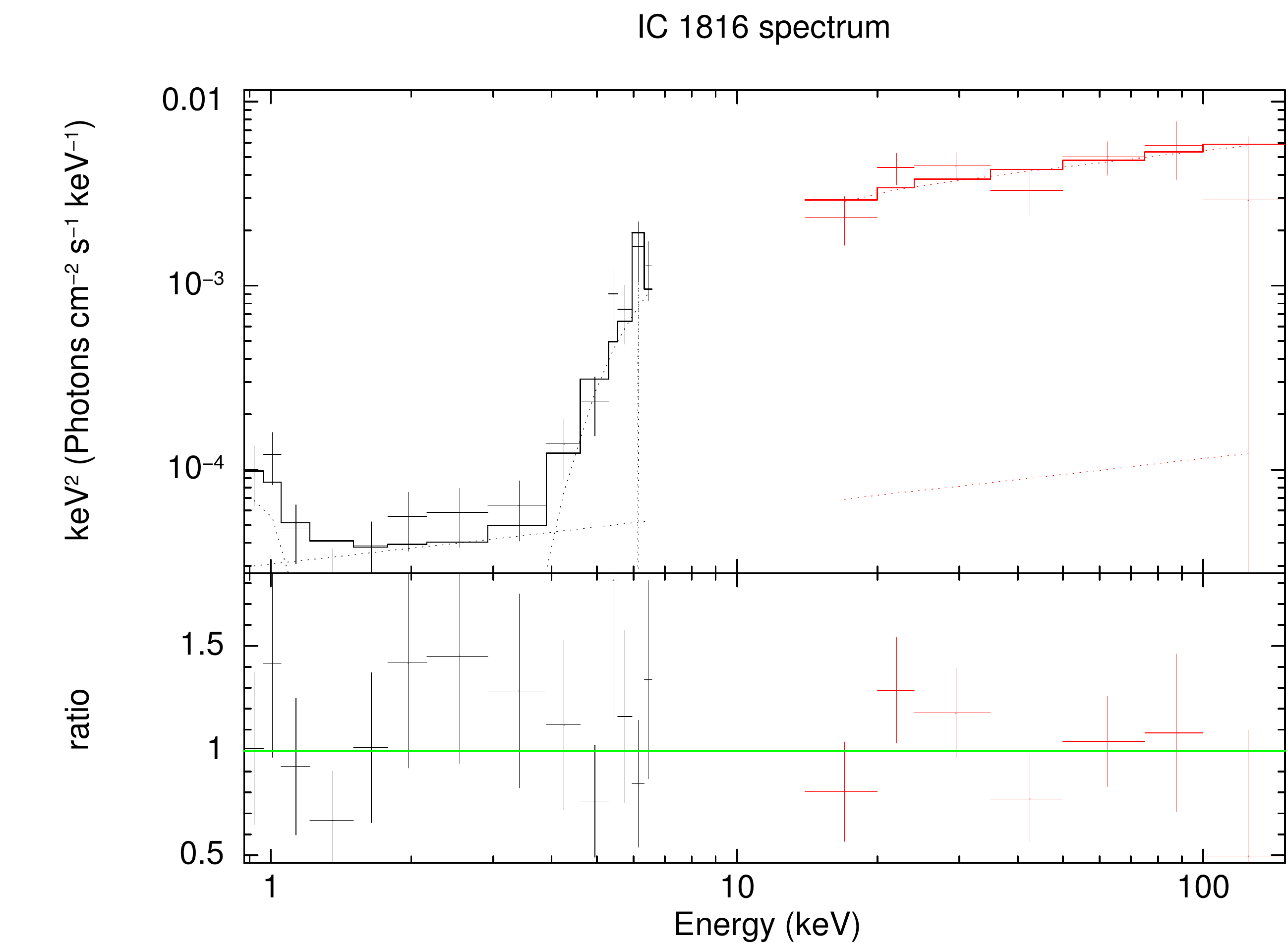}
\end{minipage}
\begin{minipage}[b]{.5\textwidth}
  \centering
  \includegraphics[angle=270,width=1.02\textwidth]{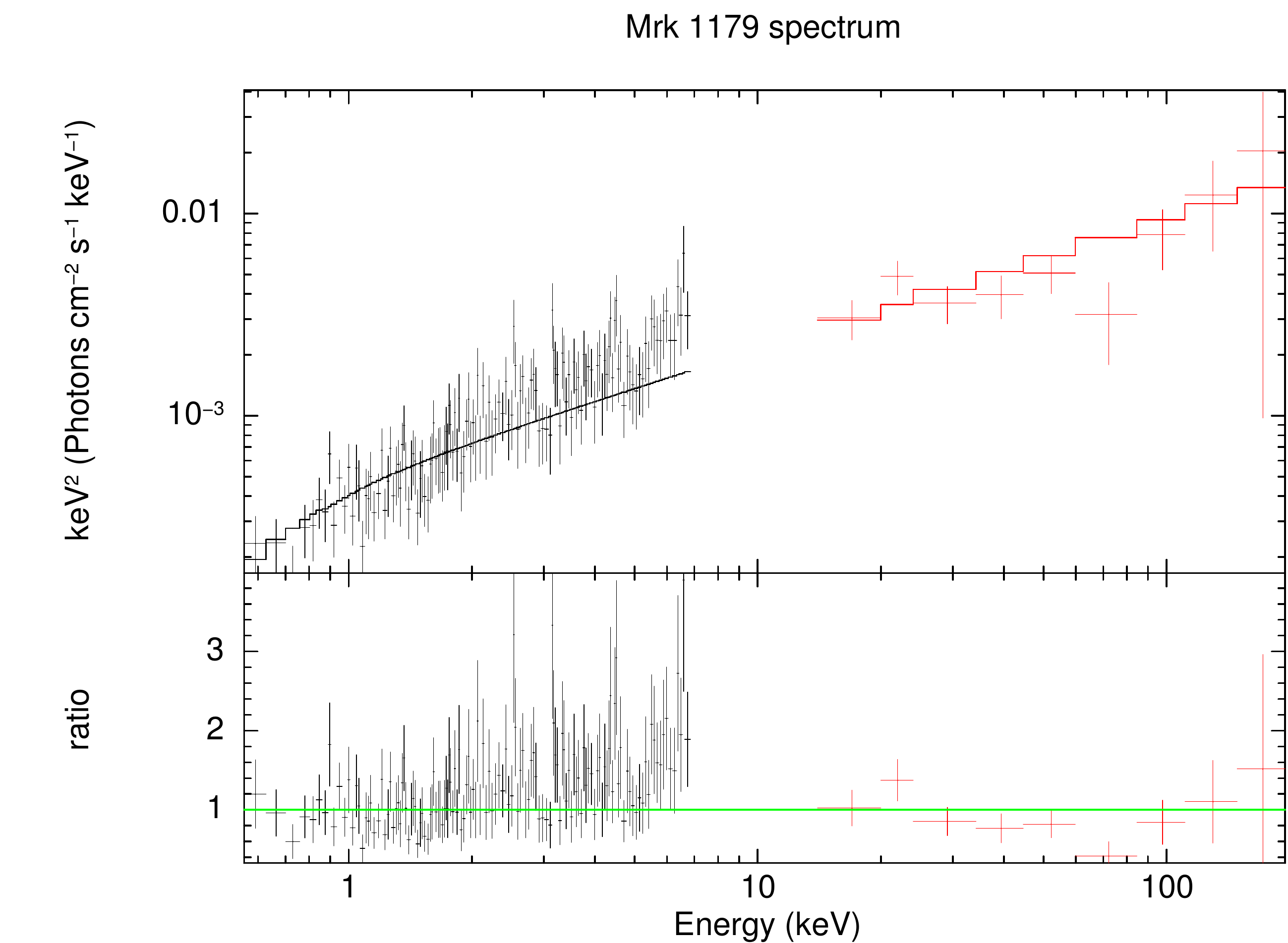}
\end{minipage}
\begin{minipage}[b]{.5\textwidth}
  \centering
  \includegraphics[angle=270,width=1.02\textwidth]{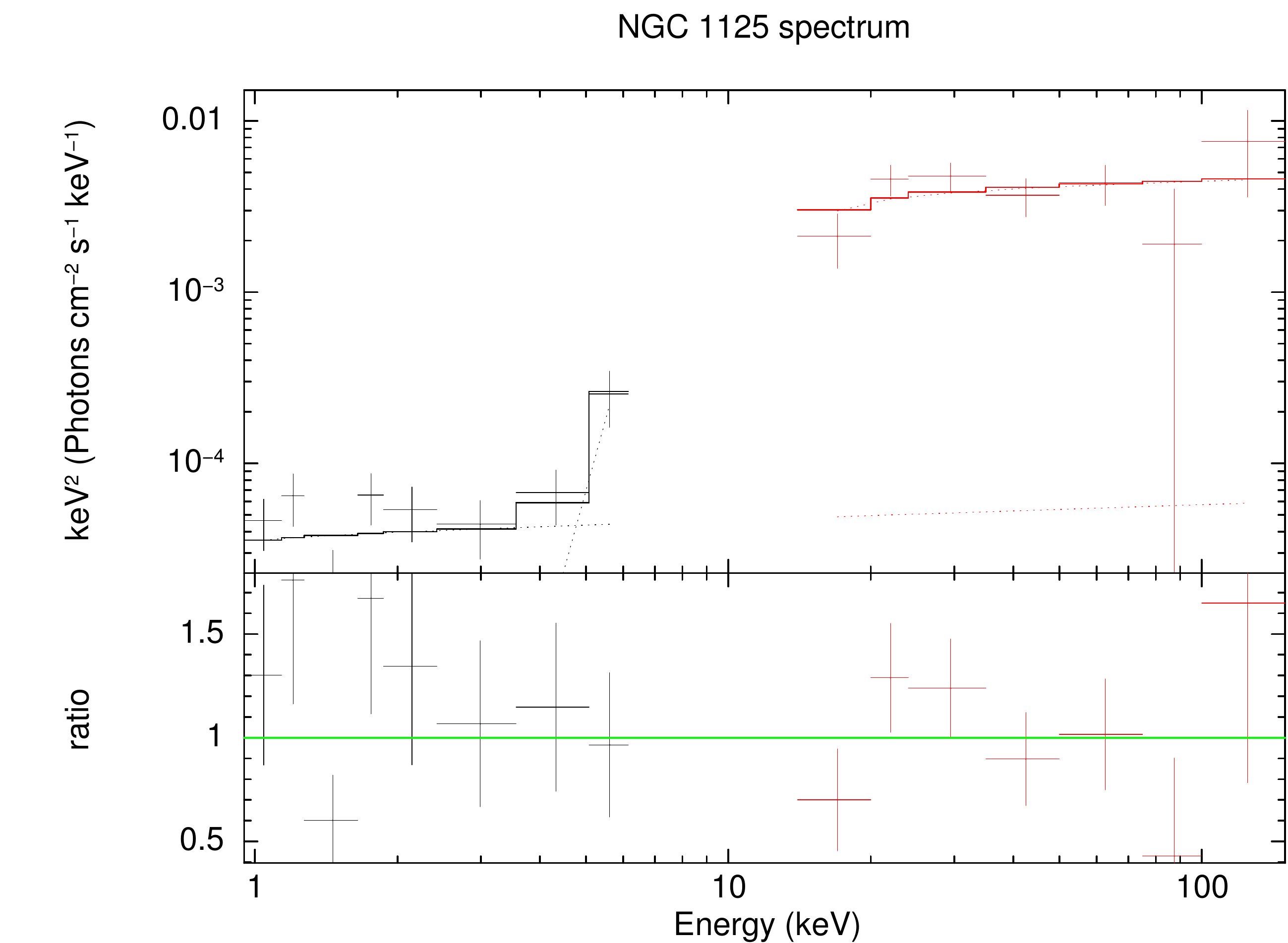}
\end{minipage}
\begin{minipage}[b]{.5\textwidth}
  \centering
  \includegraphics[angle=270,width=1.02\textwidth]{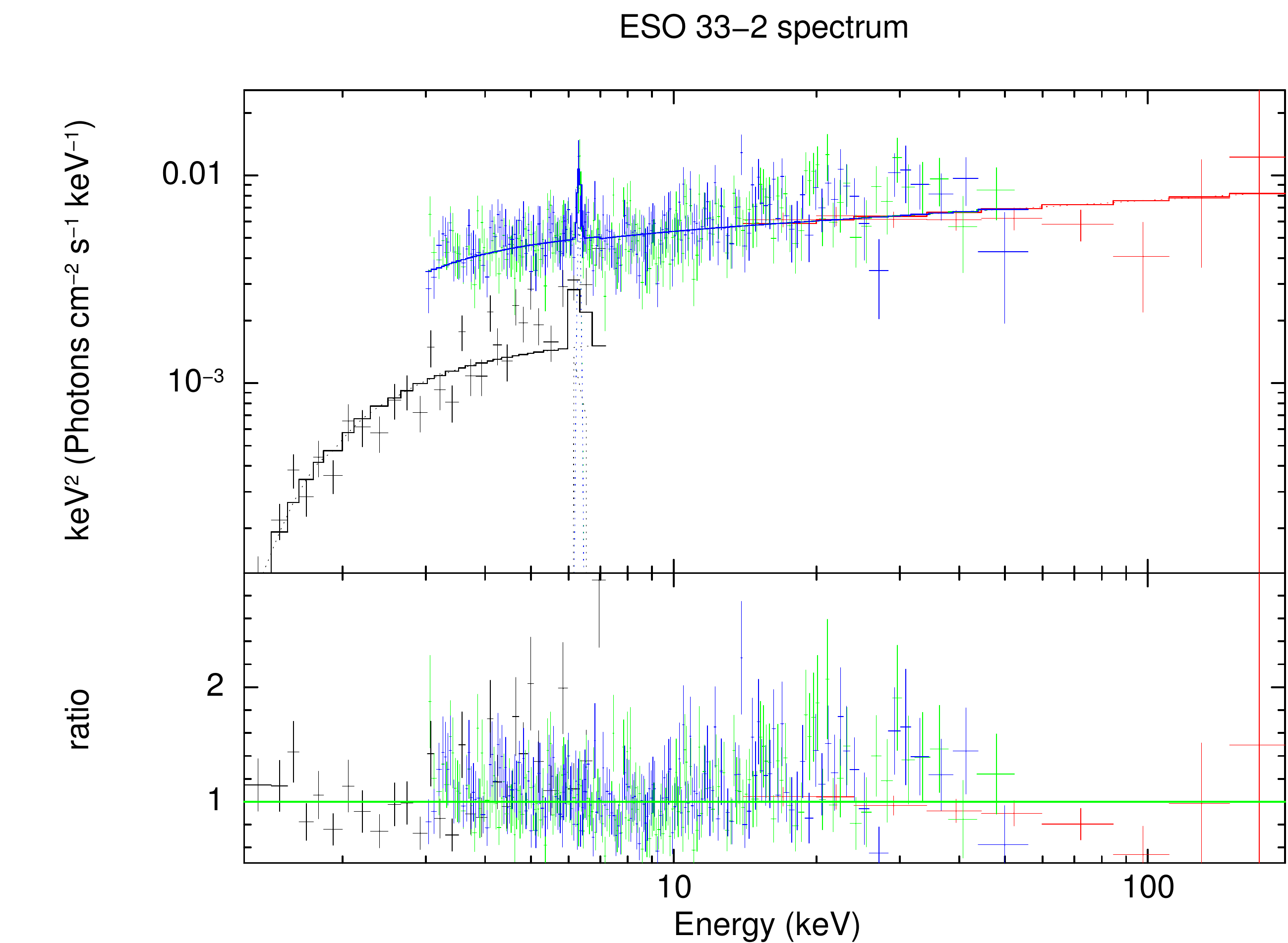}
\end{minipage}
\begin{minipage}[b]{.5\textwidth}
  \centering
  \includegraphics[angle=270,width=1.02\textwidth]{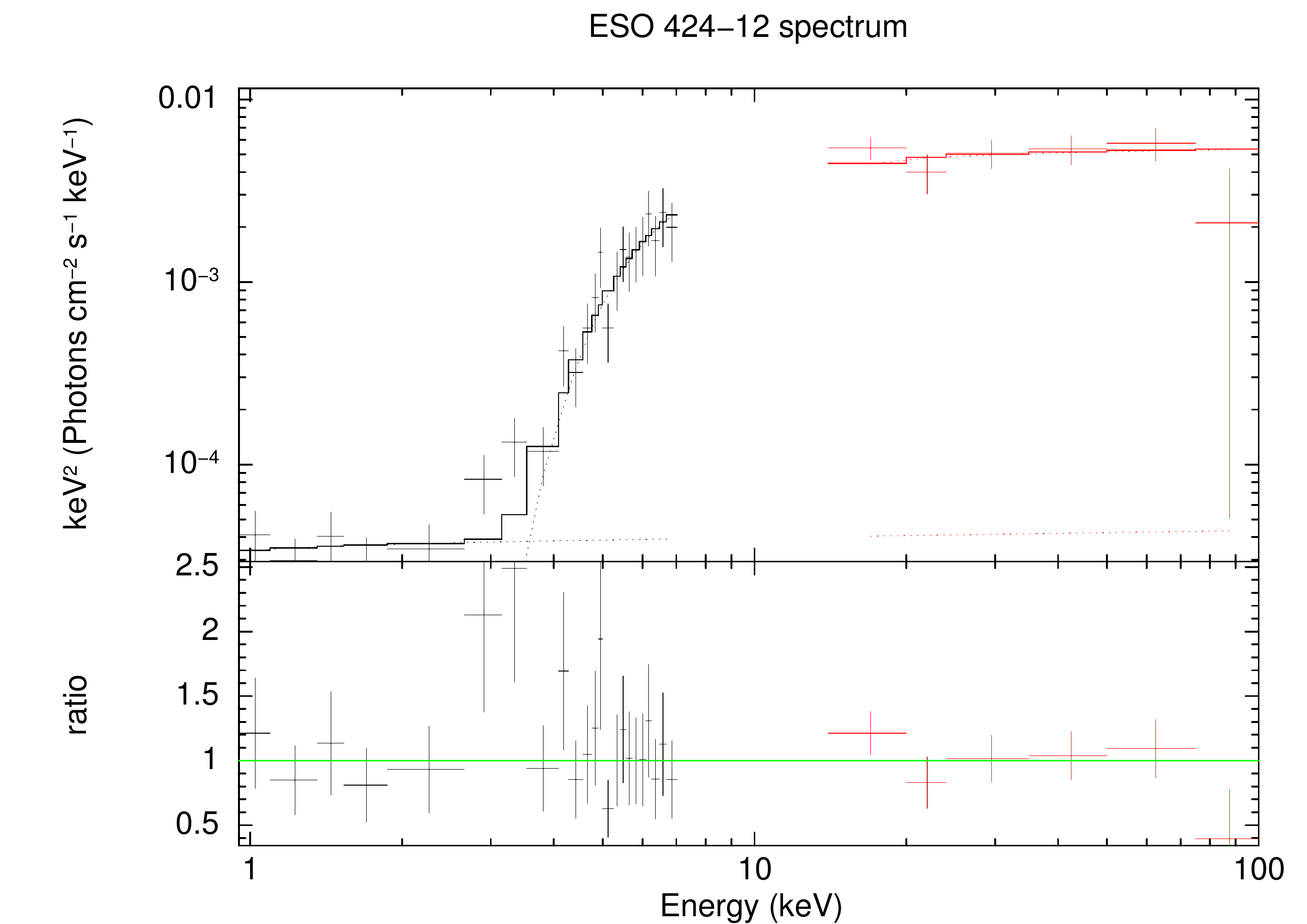}
\end{minipage}
\begin{minipage}[b]{.5\textwidth}
  \centering
  \includegraphics[angle=270,width=1.02\textwidth]{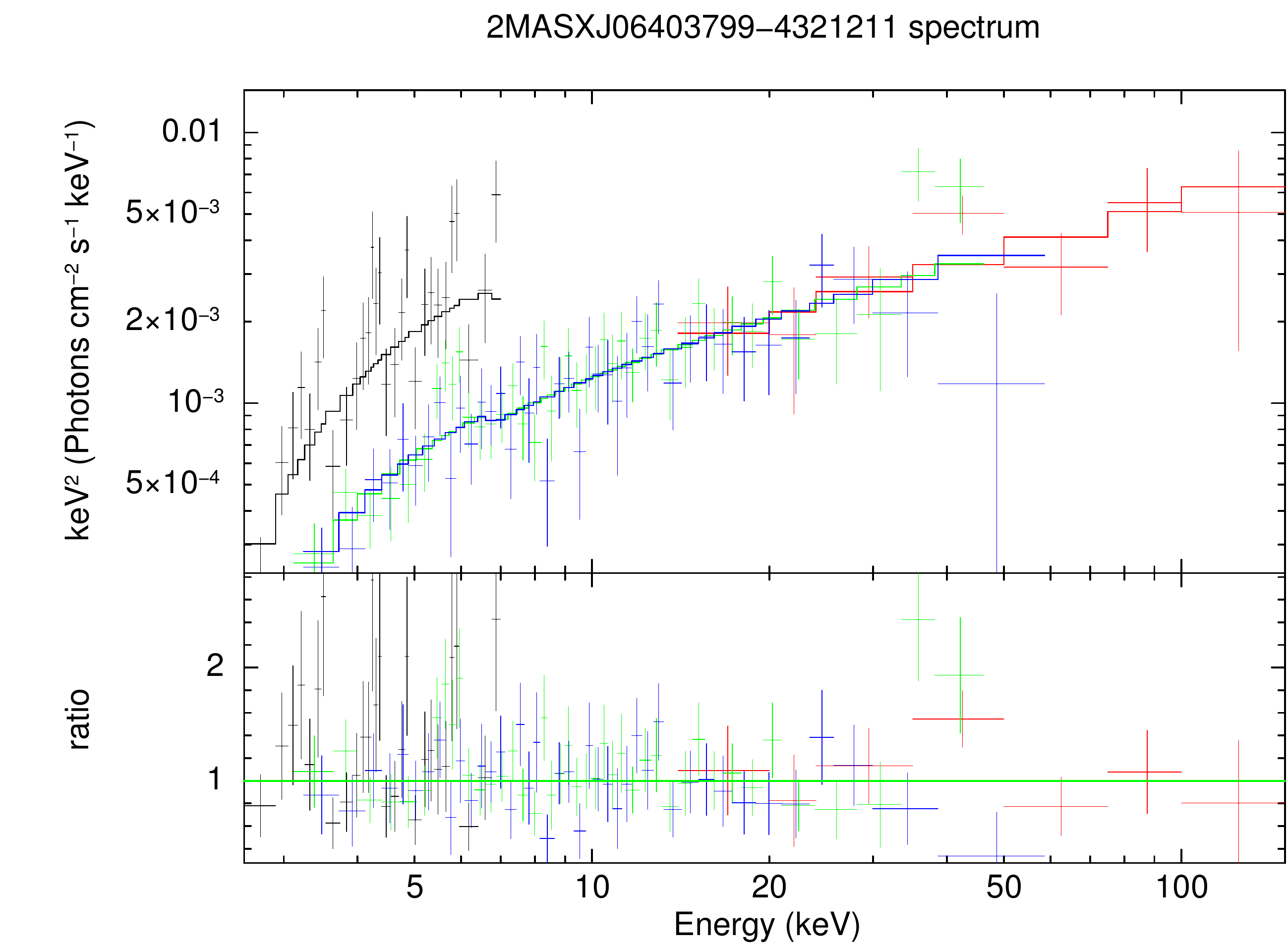}
\end{minipage}
\end{figure*}

\begin{figure*}
\begin{minipage}[b]{.5\textwidth}
  \centering
  \includegraphics[angle=270,width=1.02\textwidth]{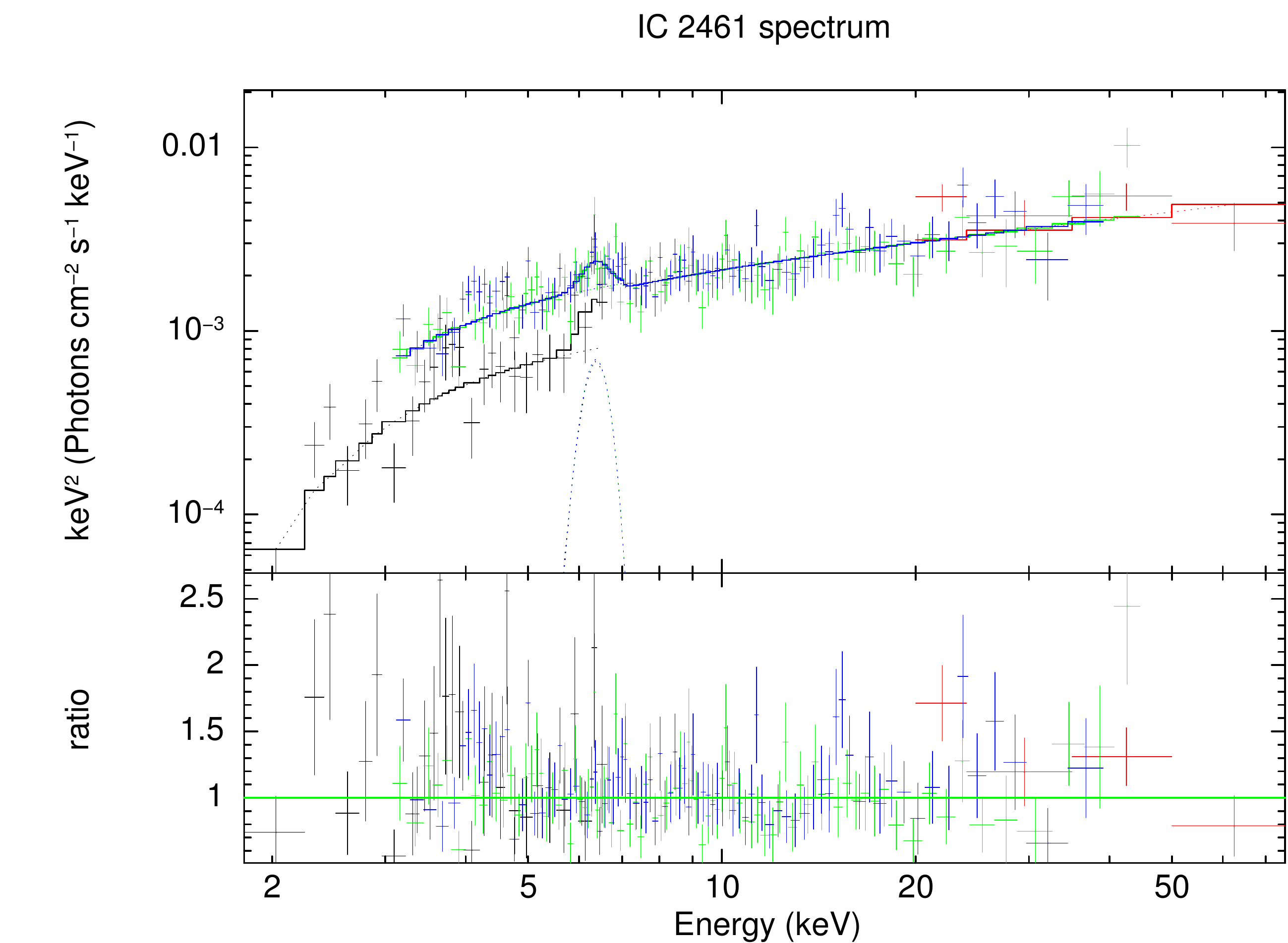}
\end{minipage}
\begin{minipage}[b]{.5\textwidth}
  \centering
  \includegraphics[angle=270,width=1.02\textwidth]{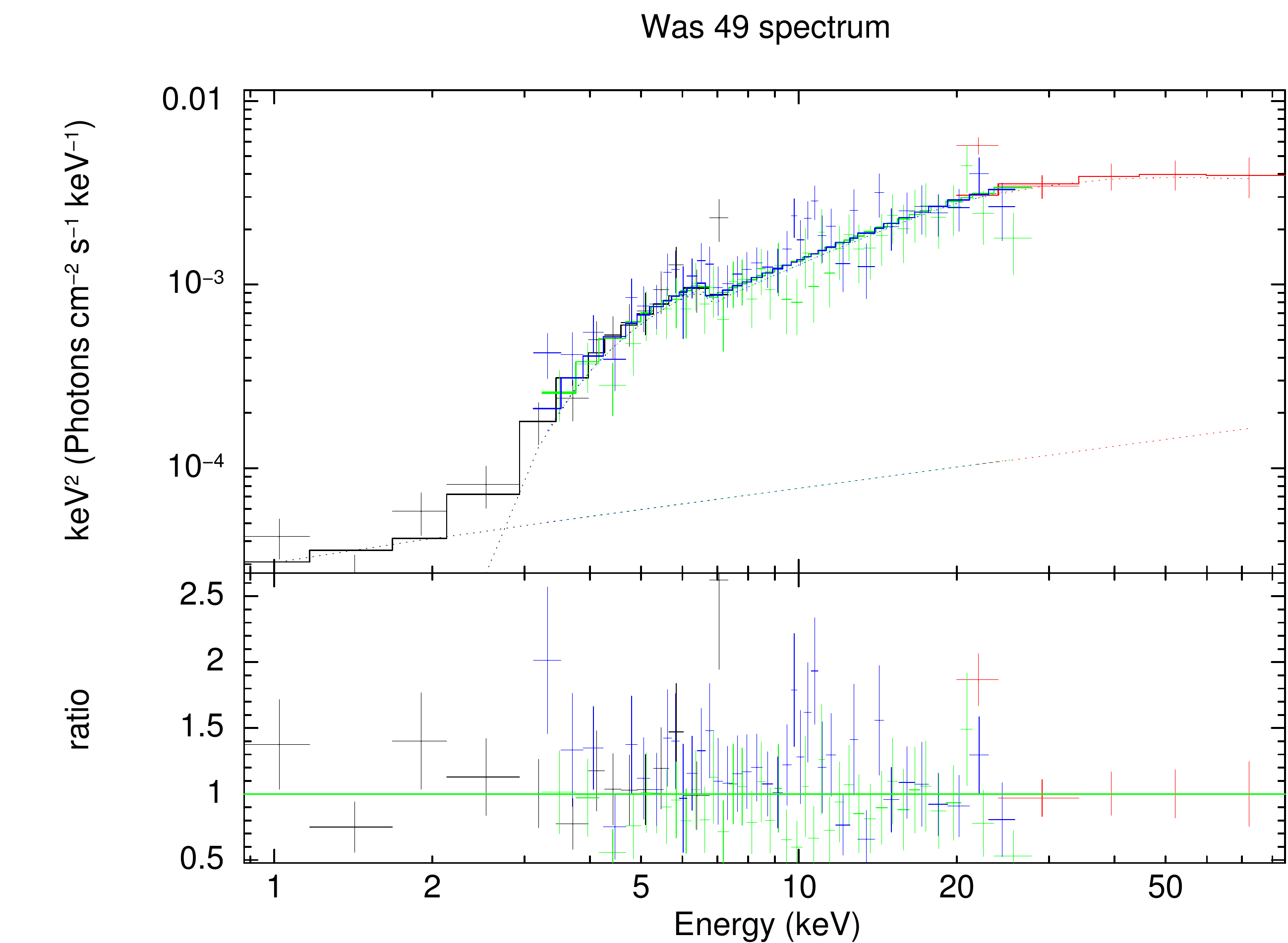}
\end{minipage}
\begin{minipage}[b]{.5\textwidth}
  \centering
  \includegraphics[angle=270,width=1.02\textwidth]{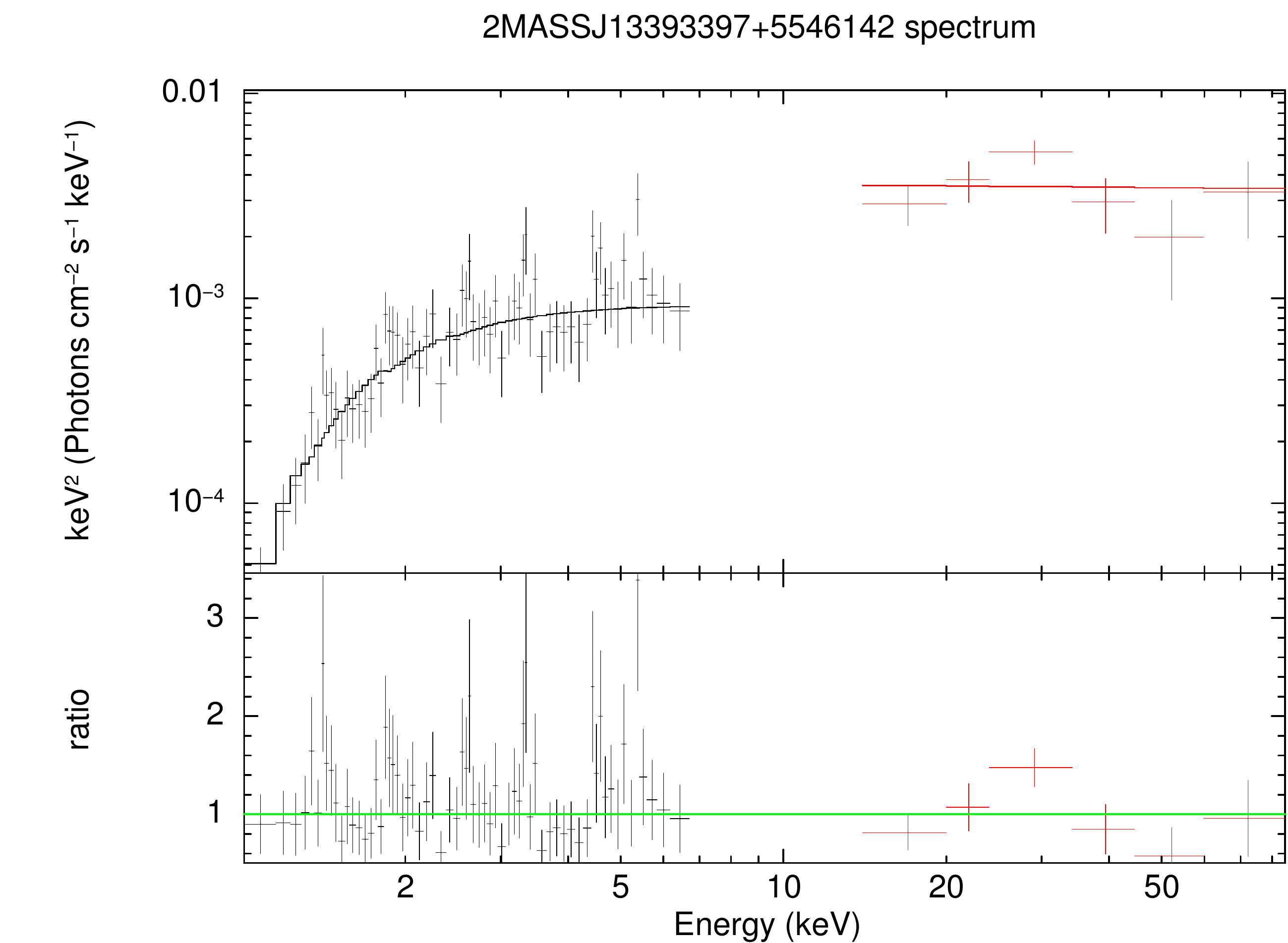}
\end{minipage}
\begin{minipage}[b]{.5\textwidth}
  \centering
  \includegraphics[angle=270,width=1.02\textwidth]{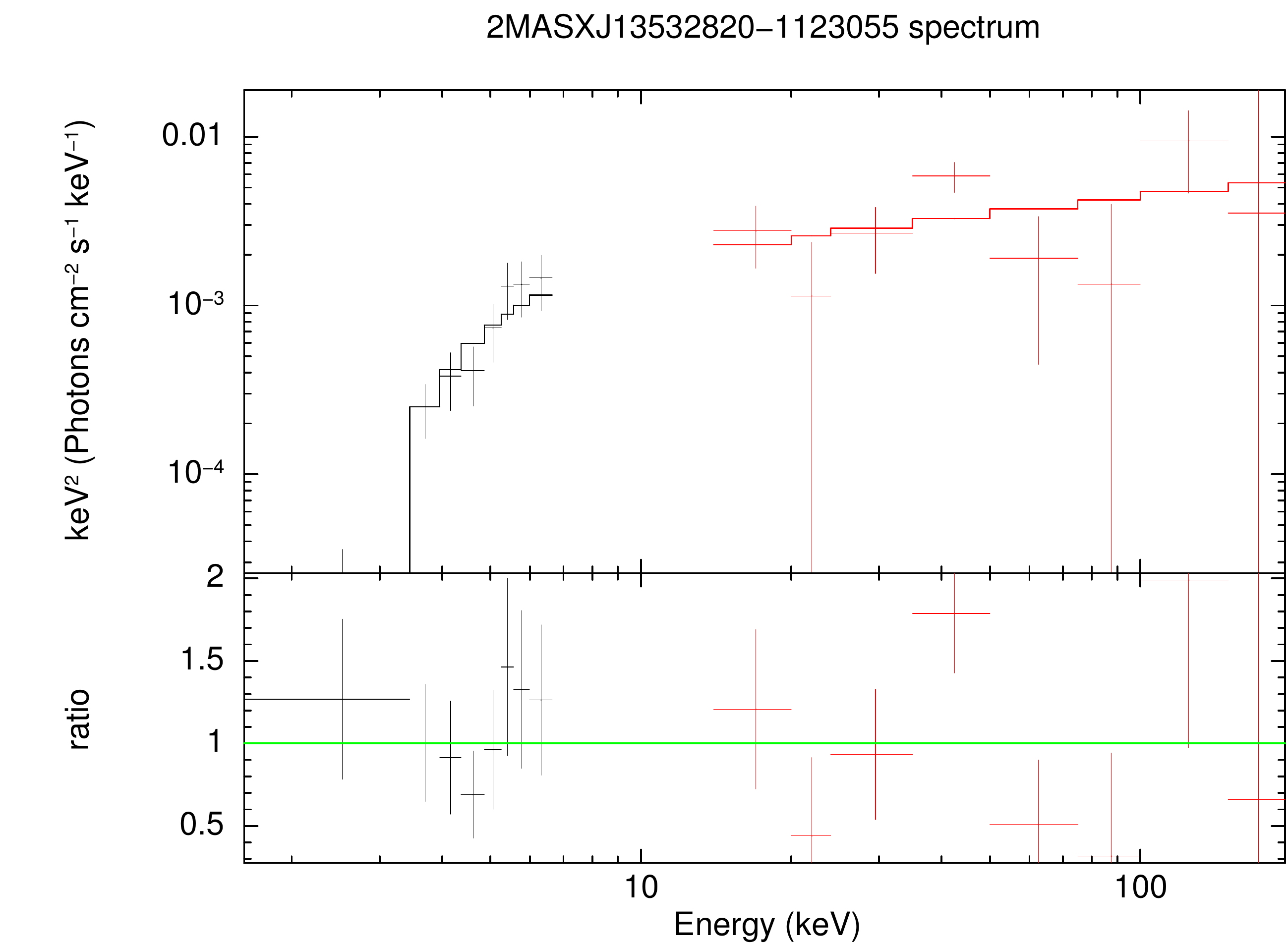}
\end{minipage}
\begin{minipage}[b]{.5\textwidth}
  \centering
  \includegraphics[angle=270,width=1.02\textwidth]{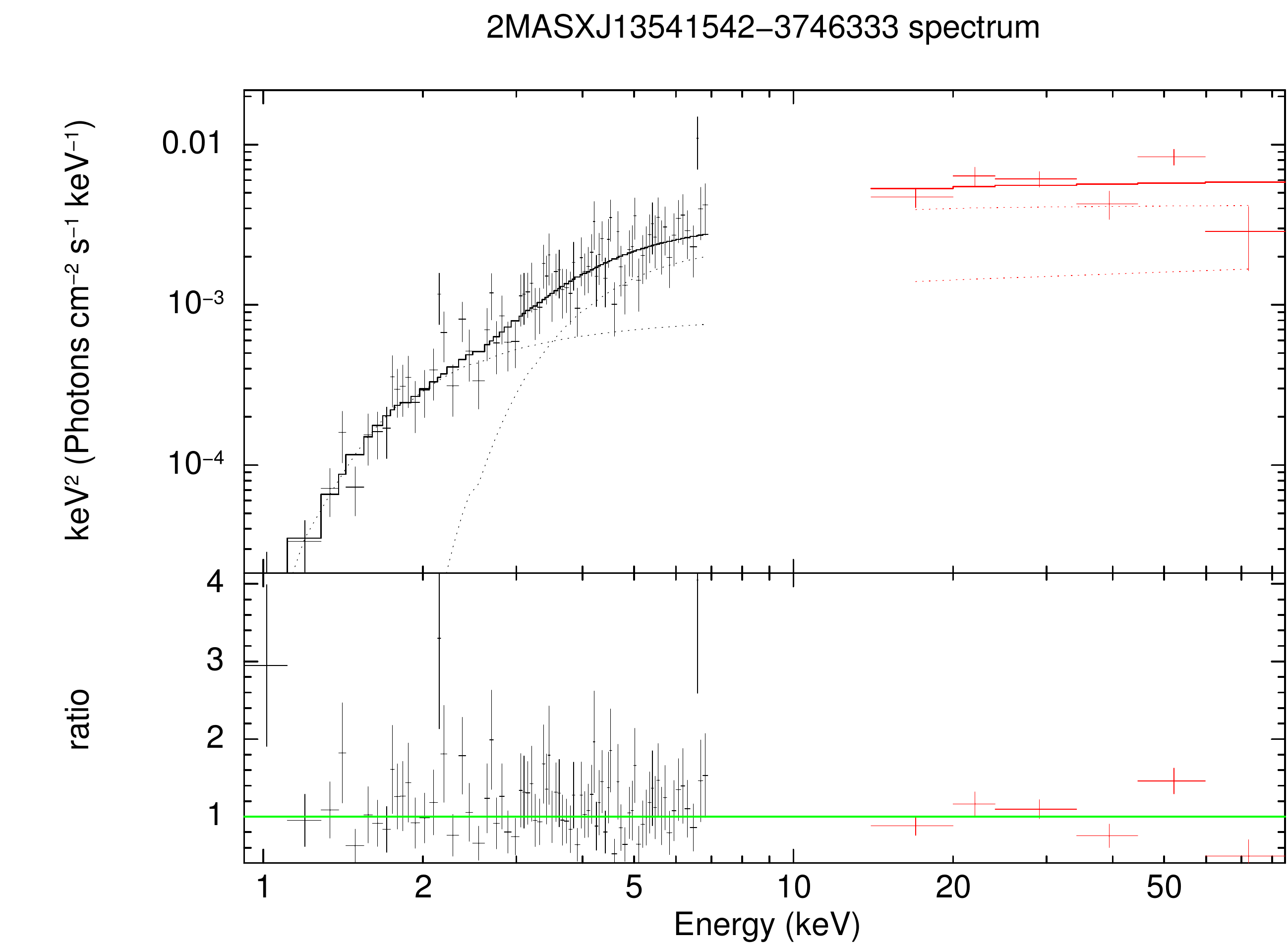}
\end{minipage}
\begin{minipage}[b]{.5\textwidth}
  \centering
  \includegraphics[angle=270,width=1.02\textwidth]{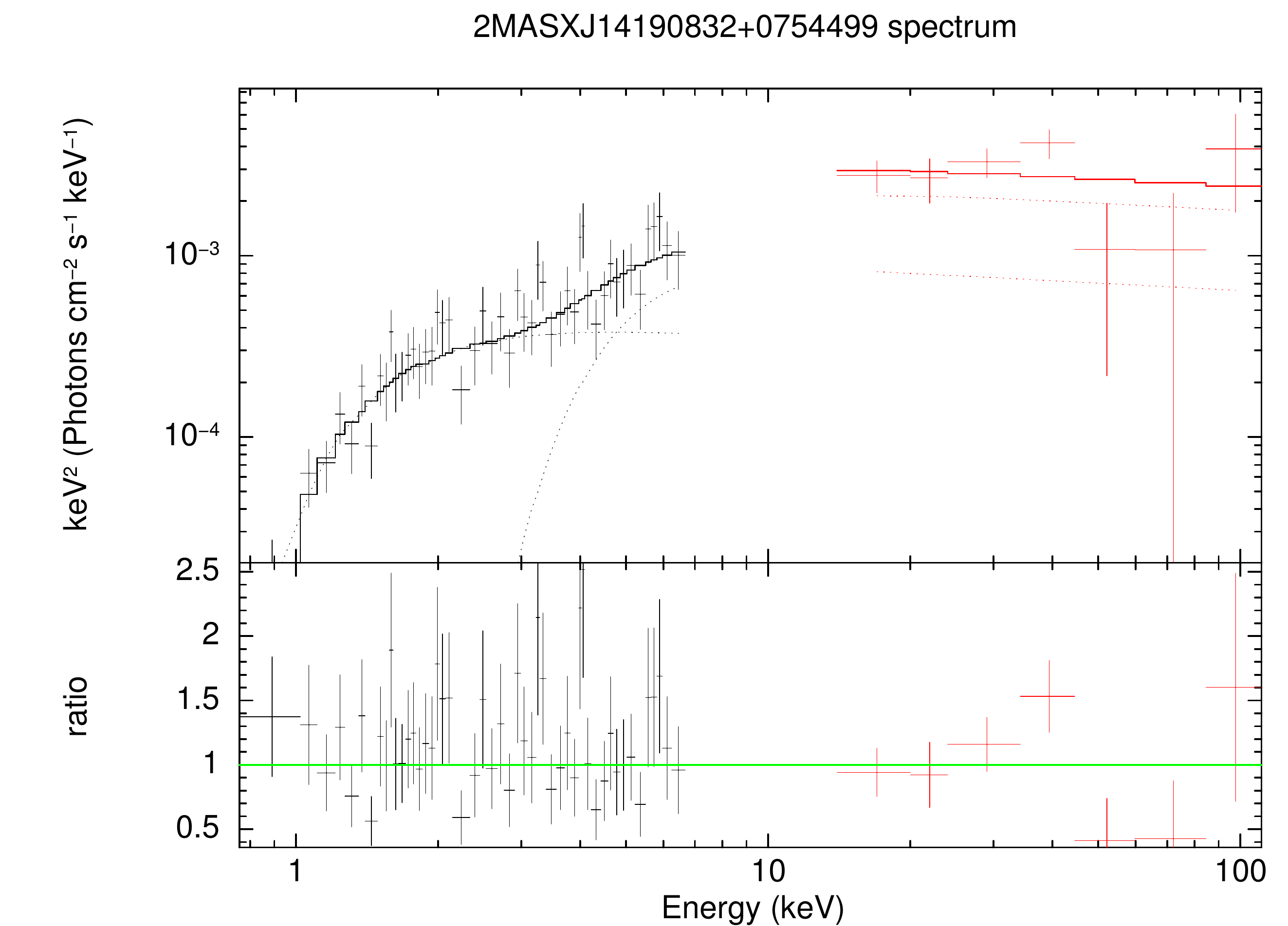}
\end{minipage}
\end{figure*}

\begin{figure*}
\begin{minipage}[b]{.5\textwidth}
  \centering
  \includegraphics[angle=270,width=1.02\textwidth]{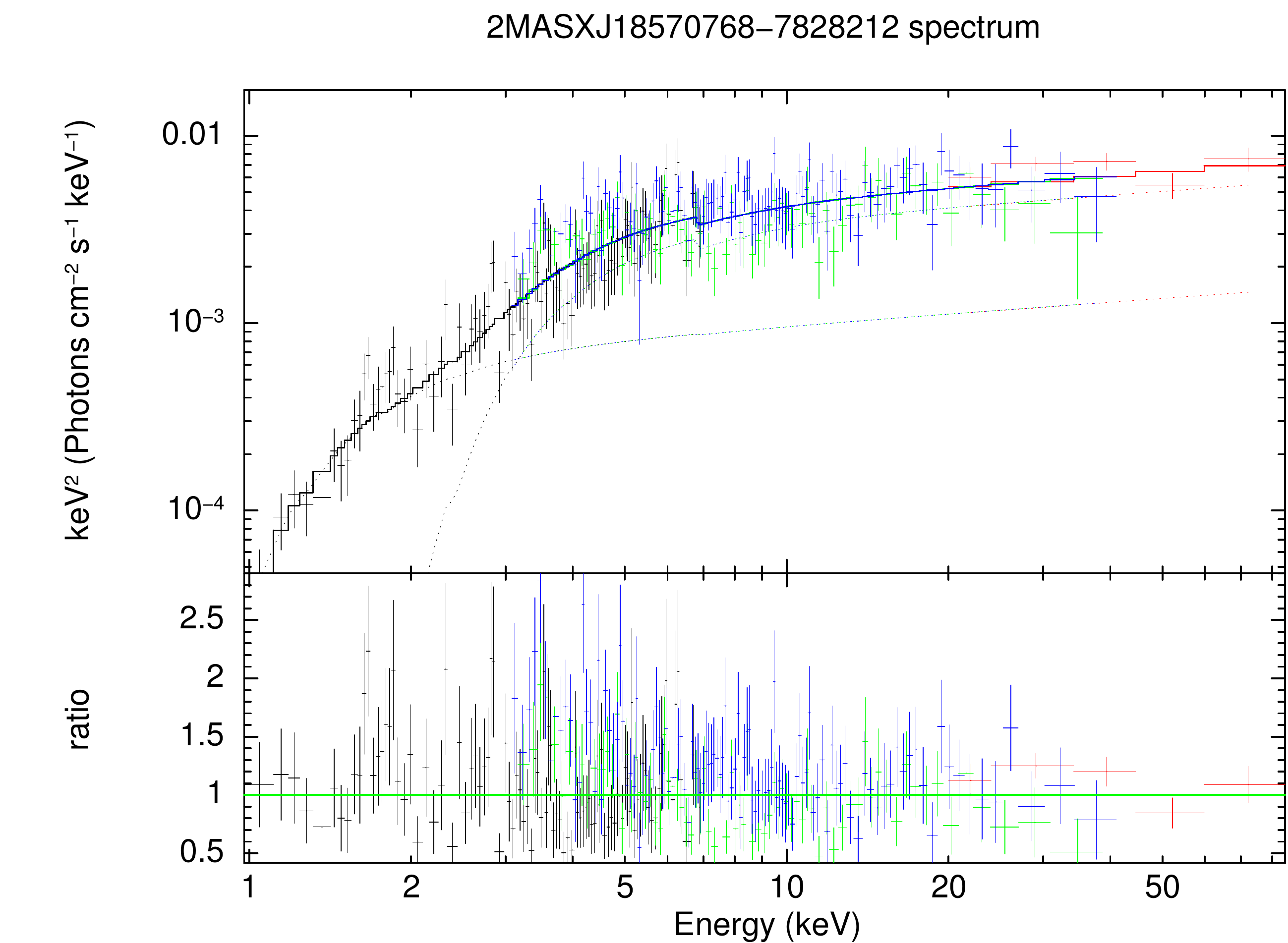}
\end{minipage}
\begin{minipage}[b]{.5\textwidth}
  \centering
  \includegraphics[angle=270,width=1.02\textwidth]{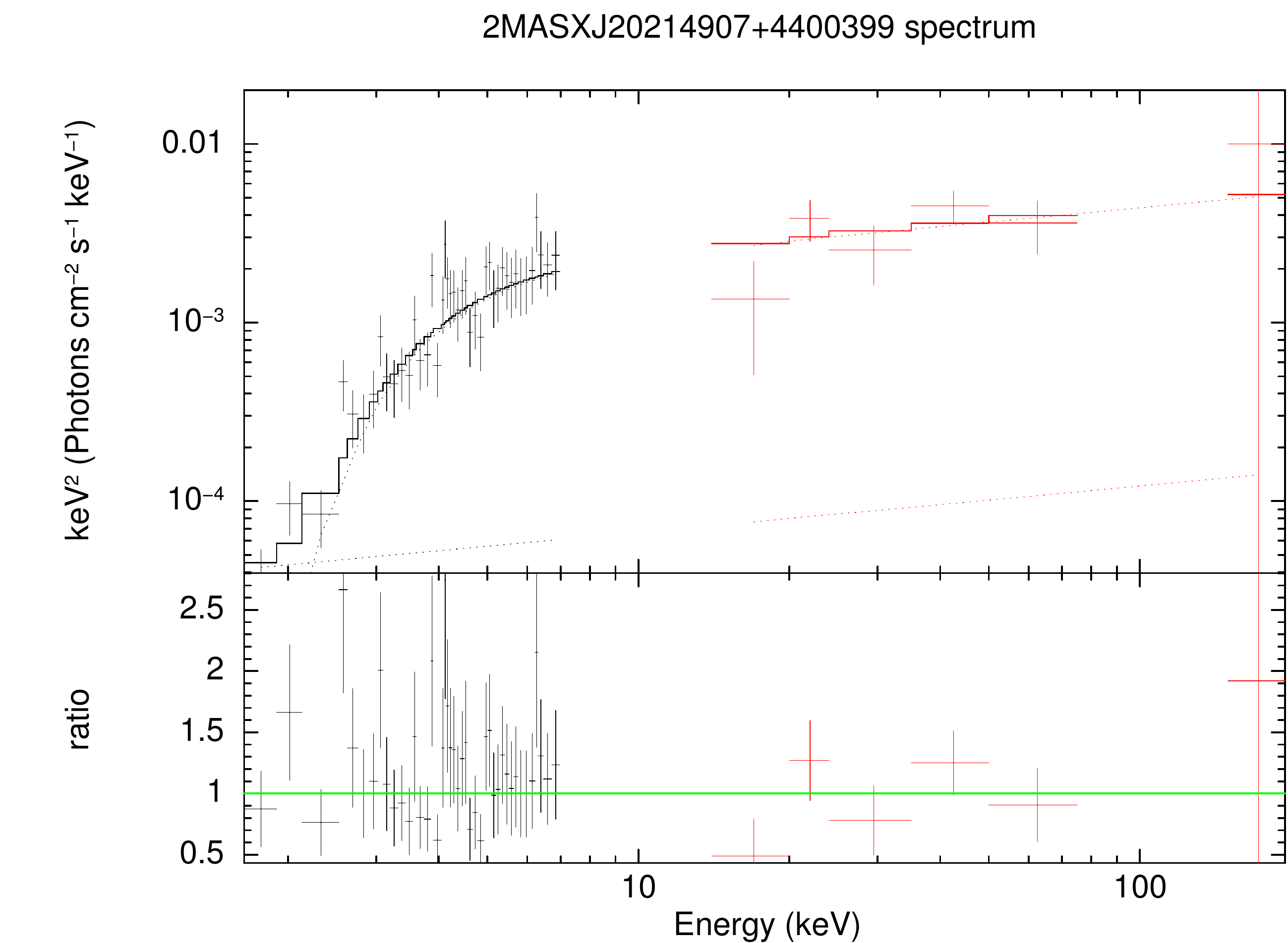}
\end{minipage}
\caption{\normalsize \cha\ (black), \nustar\ (green and blue) and \swi\ (red) spectra (top) and data-to-model ratio (bottom) of the fourteen sources in our sample.}\label{fig:spectra}
\end{figure*}

\end{appendices}

\bibliographystyle{aa}
\bibliography{BAT_chandra_follow-up}
\end{document}